\documentclass[11pt,preprintnumbers,nofootinbib,prd]{revtex4}
\usepackage{graphicx}
\usepackage{amsmath}
\usepackage{epsfig}
\usepackage{amssymb}
\newcommand{\be}{\begin{equation}}
\newcommand{\ee}{\end{equation}}
\newcommand{\ba}{\begin{eqnarray}}
\newcommand{\ea}{\end{eqnarray}}
\newcommand{\notE}{E\kern-0.6em\hbox{/}\kern0.05em}
\newcommand{\notEt}{E_{T}\kern-1.21em\hbox{/}\kern0.45em}
\newcommand{\notSUSY}{SUSY\kern-1.21em\hbox{/}\kern0.45em}
\newcommand{\GeV} {~\mathrm{GeV}}
\newcommand{\TeV} {~\mathrm{TeV}}

\def\bi{\begin{itemize}}
\def\ei{\end{itemize}}

\def\P{P_{\rm eff}}

\begin{document}

\title{{Unravelling Strings at the LHC}}
\author{{\large Gordon L. Kane}}
\author{{\large Piyush Kumar}}
\author{{\large Jing Shao}}
\affiliation{Michigan Center for Theoretical Physics, University
of Michigan, Ann Arbor, MI 48109, USA}

\vspace{2.0cm}

\date{\today}

\begin{abstract}
We construct LHC signature footprints for four semi-realistic
string/$M$ theory vacua with an MSSM visible sector. We find that
they all give rise to limited regions in LHC signature space, and
are qualitatively different from each other for understandable
reasons. We also propose a technique in which correlations of LHC
signatures can be effectively used to distinguish among these
string theory vacua. We expect the technique to be useful for more
general string vacua. We argue that further systematic analysis
with this approach will allow LHC data to disfavor or exclude
major ``corners'' of string/$M$ theory and favor others. The
technique can be used with limited integrated luminosity and
improved.

\end{abstract}
\maketitle

\newpage
\tableofcontents

\date{\today}

\section{Introduction}
The progress of string theory in the last decade has brought us
closer to the ambitious goal of connecting string theory to
reality and testing it in various experiments. However,
developments in the past few years seem to suggest that instead of
predicting a \emph{unique} well-defined vacuum from some
underlying dynamical principle, string theory gives rise to a vast
``landscape" of string vacua. From a particle physics perspective,
this implies the existence of a vast class of effective theories
for beyond-the-Standard-Model physics based on different choices
of string compactifications to four dimensions. Nevertheless, we
would like to learn more about string vacua, particularly about
aspects soon to be illuminated by LHC data.

If one is interested in connecting string theory with reality, it
is important to know if it is possible to differentiate the
effective theories arising in string compactifications from each
other based on real experimental observables, such as LHC
signatures. This question was investigated in \cite{Kane:2006yi},
where it was argued that specific string constructions usually
lead to a specific pattern of LHC signatures. In
\cite{Kane:2006yi}, the general idea and a simple method to
differentiate different classes of string constructions was
proposed. In this paper we continue our exploration along this
direction. The goals of this paper are two-fold. The first is to
demonstrate convincingly that the ``footprint" of a well-defined
class of string constructions is limited, so in particular it is
not the case that any arbitrary signature is compatible with these
``stringy" effective theories. The second is to propose a
systematic technique based on the correlation of signatures to
tell whether two classes of constructions can be distinguished or
not.


Suppose the LHC detector groups report a signal beyond the
Standard Model (SM). We expect and assume here that experimenters
and SM theorists will get that right. We want to focus on
interpreting the data in terms of an underlying theory. Most work
in this direction has tried to build a bottom-up approach by
deducing which new particles are produced, and constructing an
effective Lagrangian at the Electroweak (EW) scale. Such work
should of course be pursued. But we have increasingly learned how
difficult it may be because of issues like large number of
parameters \cite{Chung:2003fi}, degeneracies
\cite{ArkaniHamed:2005px}, etc., so complementary alternative
approaches are good.

If we knew the underlying theory at the unification scale, it
would be possible to express the many low scale effective theory
parameters in terms of perhaps a few microscopic parameters and
many degeneracies would disappear. Of course we do not know the
correct underlying theory. We argue in the following that it may
be possible to overcome this by studying a number of classes of
underlying theories and by systematically using the pattern and
correlations of LHC signatures and related data. In a sense, we
are arguing for a mapping of LHC data onto underlying theories.

Our approach can be used for any kind of underlying theory, at any
scale. We prefer to work with string/$M$ theory however, because
we expect that it will be how nature is described. Within various
string/$M$ theory models, we want to work with those which have
moduli stabilized so that reliable predictions can be made. Our
attitude is that LHC signatures and related data depend on new
particle masses and couplings and on the constraints imposed by
the underlying theory, but in a very complicated way that is
difficult to extract. By studying patterns of signatures
\cite{Kane:2006yi,Binetruy:2003cy} we can learn the implications
of the data. Insights from low scale effective theory analysis
carried out in parallel can also be included in our analysis.

Ideally, we hope there is a progress when one not only compares
different theories, e.g. Type II vs. heterotic etc., but also
takes a given type of theory and compactifies several ways, for
each compactification one can break supersymmetry several ways,
etc. One can systematically study what kinds of data can
distinguish them.

The basic idea of this paper is as follows. Based on the mapping
from model parameter space to signature space, any
Beyond-the-Standard-Model construction corresponds to a
high-dimensional sub-manifold in the signature space, which we
call the ``footprint" of the construction. Sometimes we will be
sloppy and also refer to a particular 2D slice as its footprint.
This should be clear from the context. By taking into account the
current experimental data, we may constrain the footprint. So, the
shape, size and position of a footprint carries non-trivial
information about the original construction, which is encoded in
the correlation of different signatures of the entire
construction, and also about constraints from existing data. We
develop a technique by which one could extract this information
effectively and use it to distinguish different string-theoretical
constructions. The constructions we study have already been
studied in the literature, in particular calculations for them
have been done. But for consistency we do our own calculations for
all the models.

In this paper we do not focus on details of how one scans the
microscopic parameters, their metric, SM and detector, background
and fluctuations, etc. All of these kinds of issues should be
treated in detail in application when there is data, and in a more
computer intensive study that is underway, but they do not affect
qualitative conclusions about footprints and distinguishing
theories.

\section{Realistic String Vacua}

In order to be precise, we list the criteria required for a class
of string vacua to be realistic. For concreteness, we only focus
on string vacua with low-energy supersymmetry since it appears to
be the most well motivated solution to the Hierarchy Problem;
however realistic string vacua with other methods of explaining
the Hierarchy can be similarly defined.
\subsection{String-Susy Models}
To qualify as what we call a ``String-Susy Model", we require a
class of string vacua arising from a compactification to four
dimensions to have the following properties:
\begin{itemize}
\item It has ${\cal N}=1$ supersymmetry in four dimensions which
is broken in a controlled approximation. \item The moduli are
stabilized in a metastable dS vacuum and a stable hierarchy
between the Electroweak and Planck Scales is generated. \item The
visible sector accommodates the MSSM particle content and gauge
group (maybe with additional matter and gauge groups) and their
properties. \item It has a mechanism for breaking the Electroweak
symmetry.\item It is consistent with all experimental constraints.
\end{itemize}
In addition, the fact that gauge couplings in the MSSM unify with
great precision at $\sim 2\times10^{16}$ GeV seems to be
tantalizing evidence for gauge coupling unification and a unified
theory framework. Although some string-susy models give rise to
gauge coupling unification naturally and some don't (for example,
LARGE Volume models do not give rise to gauge coupling unification
with an MSSM visible sector), gauge coupling unification is still
an important criterion in our opinion and should serve as an
important guide for constructing realistic string theory vacua.

Present models do not quite meet these criteria, but are close
enough to justify working with them. More precisely, we study four
dimensional vacua in string theory where the first two conditions
are met in a reliable manner, i.e. the question of supersymmetry
breaking, moduli stabilization and generation of the Hierarchy is
answered in a convincing manner. One popular example is KKLT vacua
proposed by Kachru, Kallosh, Linde and Trivedi \cite{KKLT1}. The
KKLT vacua do not comprise a string-susy model in the strict sense
since an explicit compactification using the KKLT mechanism to
stabilize the moduli and containing a visible sector satisfying
the last two conditions has not yet been constructed. However,
KKLT vacua still allow a ``prediction" if a certain visible sector
particle content is assumed. For example, if the particle content
is assumed to be that of the MSSM, it picks out a subset of MSSM
models which encodes features of KKLT vacua in their spectra and
signature pattern. This procedure can be carried out for other
classes of vacua as well. One thus obtains classes of models, each
of which are completely specified by a set of properly chosen
stringy/microscopic parameters characterizing the particular class
of string vacua. By a slight abuse of notation, we will still call
them ``string-susy models". The consequences of such MSSM models
at the LHC can be readily predicted by standard methods. For
simplicity and concreteness, in this paper we assume that an MSSM
visible sector particle content is realized for each string-susy
model. It would be very interesting to relax this requirement in
the future and study the consequences. When concrete matter
embeddings are available in the above classes of models, there may
arise extensions of the MSSM. It will be interesting to study
them.

In this paper, we study three well-motivated classes of string
vacua assuming an MSSM visible sector - Type IIB KKLT vacua
\cite{KKLT1,KKLT2}, Type IIB LARGE-Volume vacua \cite{LargeVolume}
and fluxless $M$ theory vacua \cite{Acharya:2006ia,Acharya:2007rc}
on $G_2$ manifolds. For KKLT vacua, we also study a variation of
the original KKLT procedure which uses the mechanism of $F$-term
uplifting rather than that by anti D3-branes
\cite{Lebedev:2006qq}-\cite{Lebedev:2006qc}. Each of these vacua
have been studied in the literature in detail and the interested
reader can consult the relevant references. Next we briefly
summarize their most important features.

If an MSSM visible content is realized in these constructions, a
set of soft supersymmetry breaking parameters will be generated
after supersymmetry breaking. These MSSMs together with a set of
soft supersymmetry breaking parameters then constitute our
``string-susy MSSMs".

\subsection{Description of String-Susy Models}

\subsubsection{(Original) KKLT MSSM vacua - SUSY breaking by ${\overline D3}$-branes (KKLT-1)}\label{Sec:KKLT}

This class of constructions is a part of the IIB landscape with
all moduli stabilized \cite{KKLT1}. Closed string fluxes are used
to stabilize the dilaton and complex structure moduli at a high
scale and non-perturbative corrections to the superpotential are
used to stabilize the lighter K\"{a}hler moduli. One obtains a
supersymmetric anti-deSitter vacuum. The hidden sector is an
anti-D3-brane at the IR end of the throat, and is thus sequestered
from the visible sector. The anti D3-brane breaks supersymmetry as
well as lifts the vacuum to a deSitter one. Supersymmetry breaking
is then mediated to the visible sector by gravity. The flux
superpotential ($W_0$) has to be tuned very small to get a
gravitino mass of $\mathcal{O}$(1-10 TeV). The soft supersymmetry
breaking terms at the unification scale are calculated in
\cite{KKLT2}:
\begin{eqnarray}
  M_a&=&M_s\;\Big[l_a\alpha+b_{\alpha}g_{a}^2\Big],\nonumber\\
  m_i^2&=&M_s^2\Big[(1-n_i) \alpha^2+4\alpha \xi_i-\dot\gamma_i\Big],\nonumber\\
  A_{ijk}&=&-M_s\Big[(3-n_i-n_j-n_k)\alpha-\gamma_i-\gamma_j-\gamma_k\Big], \label{KKLTsoft}
\end{eqnarray}
where $b_a$ are the $\beta$ function coefficients, $\gamma_i$ is
the anomalous dimension and ${\dot \gamma}_i=8\pi^2\frac{\partial
\gamma_i}{\partial \ln\mu}$. The coefficient $\xi_i$ is a
complicated function of trilinear couplings, Yukawa couplings and
gauge couplings \cite{KKLT2}. Here $M_s\equiv m_{3/2}/(16\pi^2)$
characterizes the size of the AMSB contribution and $\alpha$ is
the ratio of the modulus-mediated contribution to the AMSB
contribution, defined as in \cite{Baer:2006id} \footnote{Note this
definition is different from that in \cite{Choi:2007ka}.
}. The parameter $\alpha$ is determined by the form of the
uplifting potential and the flux contribution. In
\cite{Choi:2007ka}, it was argued that typically $\alpha$ can take
a generic value of order unity, which in the definition of
\cite{Baer:2006id} is of order $16\pi^2/\ln(M_p/m_{3/2})\sim 5$
for $m_{3/2}=(1-10)\TeV$.

The SM gauge fields can live on D7-branes or D3-branes, which
corresponds to $l_a=1 \; \rm {or}\;0$ respectively. We will focus
on the former case and set $l_a=1$. The chiral matter fields can
be constructed by adding intersecting D7-branes with magnetic
fluxes in their worldvolume. In the case of toroidal (orbifold)
compactifications and no magnetic fluxes, the modular weights
$n_i$ can take values $0$, $1/2$ or $1$ depending on whether the
matter fields are on the D7-brane, D3-D7 intersection or D3-brane
respectively. For compactifications with more general Calabi-Yau
manifolds or with more general intersecting D7-brane models with
worldvolume magnetic fluxes, $n_i$ will be model dependent and
have to be computed in each model\cite{Ibanez:2004iv}. Generally
if the modular weighs are equal to 1, all the scalars will be
tachyonic as a result of the mixing of the moduli and AMSB
contribution in Eq.(\ref{KKLTsoft}). So modular weights of 1 are
normally excluded. In addition, the ratios of gaugino masses at
low scale are roughly $(1+3.3/\alpha):(2+1/\alpha):(6-9/\alpha)$.
For a typical value of $\alpha=5$, the ratio is
$1.5\,:\,2\,:\,3.8$. The LSP is predominantly bino-like for a
sizable range of $\alpha$ around $5$. The constraint on the relic
density of neutralino dark matter favors the region in the
parameter space where there is some bino-wino mixing (not
necessarily large), or, a stop or stau with mass close to that of
LSP.


\subsubsection{KKLT MSSM vacua - SUSY breaking by hidden sector
$F$-terms (KKLT-2)}\label{Sec:KKLT-2}

There is a variation of the original KKLT proposal in which the
anti D3-brane is replaced by a hidden sector which spontaneously
breaks supersymmetry and lifts the AdS minimum. This is also known
as $F$-term uplifting. Several examples of this type of vacua are
discussed in the literature
\cite{Lebedev:2006qq}-\cite{Lebedev:2006qc}. A notable example is
to use the recently discovered ISS model
\cite{Intriligator:2006dd} as the hidden section, which can
potentially have a dual stringy construction via AdS/CFT duality
\cite{Argurio:2007qk}. In this example, dS vacua with zero
cosmological constant and TeV scale gravitino mass can both be
realized naturally at the same time \cite{Dudas:2006gr}.

The phenomenology associated with this class of vacua is
model-dependent and is still under investigation. To our
knowledge, a generic parametrization of soft supersymmetry
breaking terms can be found in \cite{Lebedev:2006qc} and will be
used in our analysis. In this result, the gaugino masses and
trilinears are similar to those in the original KKLT proposal,
while the scalar masses are of the form:
\begin{eqnarray}
  m_i^2=(16\pi^2 M_s)^2(1-3\zeta_i).
\end{eqnarray}
Here $\zeta_i$ are the couplings entering the matter K\"{a}hler
potential:
\begin{eqnarray}
  K_{\rm matter}\sim \bar Q_i Q_i(T+\bar T)^{n_i}\left[1+\zeta_i\bar\phi\phi+{\cal
  O}(\phi^4)\right].
\end{eqnarray}
where $Q_i$ are the visible sector matter fields and $\phi$ is the
hidden sector matter field. In general without special assumption
on the construction (e.g. geometric separation), the hidden sector
is not sequestered from the visible sector. So $\zeta_i$ are
expected to be of order unity, which gives rise to unsuppressed
scalar masses ($\sim m_{3/2}$). Thus the modular weights $n_i$ are
not important in determining soft terms and are set to zero for
convenience.
In the limit $\zeta_i\rightarrow 1/3$, the scalars become light
and the mirage pattern of scalar masses is recovered. In addition,
the non-sequestering of the hidden sector also implies a larger
range of values of $\alpha$ \cite{Lebedev:2006qc}. In our
phenomenological analysis, this class of models will be referred
to as KKLT-2.

\subsubsection{LARGE Volume MSSM vacua (LGVol)}\label{Sec:LGVOL}

This class of constructions also form part of the IIB landscape
with all moduli stabilized. In this case, the internal manifold
admits a large volume limit with the overall volume modulus very
large\footnote{We distinguish between what is usually called large
volume, where manifold volumes are several times the volume in
Planck units, and the volumes of manifolds for these models where
the volume is several orders of magnitude larger than the volume
in Planck units, giving rise to an intermediate scale string
scale. We denote the latter case by LARGE volume.} and all the
remaining moduli small \cite{LargeVolume}. Fluxes again stabilize
the complex structure and dilaton moduli at a high scale, but the
flux superpotential $W_0$ in this case can be $\mathcal{O}$(1). To
stabilize K\"{a}hler moduli in the large volume region, one needs
to incorporate the perturbative contributions ($\alpha'$
correction) to the K\"{a}hler potential as well as the
non-perturbative contributions to the superpotential since they
are equally important. The AdS minimum of the resulting potential
is already non-supersymmetric in contrast to the KKLT case, which
can be lifted to a de Sitter one by similar mechanisms as in the
KKLT case.

This class of vacua turns out to be more general and includes the
KKLT vacua as a special limit, in which $W_0$ is tuned very small
\cite{LargeVolume,Quevedo06}. However, when $W_0$ is
$\mathcal{O}$(1), the conclusions are qualitatively different. We
will analyze such a situation, since then there will be no
theoretical overlap between the two classes of vacua. The
exponentially large volume $\cal V$ generated allows both lowered
string scale and gravitino mass
\begin{eqnarray}
  m_s\sim \frac{M_P}{\sqrt{{\cal V}}}, \quad m_{3/2}\sim
  \frac{M_P}{{\cal V}}.
\end{eqnarray}
To get a TeV-scale gravitino mass, one needs a volume ${\cal
V}\sim 10^{15}$ which gives rise to an intermediate string scale
$m_s\sim 10^{11}$ GeV. Since the string scale is much smaller than
the unification scale, one cannot have the standard gauge
unification in these compactifications with only MSSM matter.

In this class of constructions, the Standard Model sector arises
from an appropriate configuration of D7-branes which wrap a small
four-cycle (a four dimensional submanifold of the entire
Calabi-Yau manifold) corresponding to the modulus $\tau_s$. To
generate chirality, the SM D7-brane is required to be magnetized,
which gives rise to a modified gauge kinetic function
\begin{eqnarray}
  f_i=\frac{T_i}{4\pi}+h_i(F)S,
\end{eqnarray}
where $h_i$ is a topological function of the magnetic flux on the
brane.
An exact calculation of the gaugino masses at the lowered string
scale gives the following boundary condition
\begin{eqnarray}
  M_1:M_2:M_3=k_Yg_1^2:g_2^2:g_3^2,
\end{eqnarray}
where $k_Y$ is determined by the normalization of the U(1) charge.
Since there is no gauge coupling unification in this construction,
the gaugino masses are also not unified at the string scale.

The scale of the gaugino masses is determined by the F-term $F^s$
of the small four-cycle, which is characterized by
\begin{eqnarray}
M_c\equiv\frac{F^s}{2\tau_{s}}\approx \frac{1}{2}(M_2+M_3),
\end{eqnarray}
where $\tau_s={\rm Re}(T_s)$ is the modulus associated with the
small four-cycle. There is a so-called ``dilute flux limit" in
which the magnetic flux is diluted by increasing the size of the
large four-cycle. In such a case, it was shown in
\cite{Conlon:2007xv} that the scalar and trilinear terms at high
scale take simple expressions and are given by:
\begin{eqnarray}
  m_i=\frac{1}{\sqrt{3}}\frac{F^s}{2\tau_s}=\frac{M_c}{\sqrt{3}}\\
  A_{ijk}=-\frac{F^s}{2\tau_s}=-M_c.
\end{eqnarray}
Generally the presence of the fluxes will modify the above
equations. The effects of these fluxes are modelled by small
perturbations $\epsilon_i$ around the above results. Therefore the
LARGE Volume soft spectrum can be parameterized as in
\cite{Conlon:2007xv}
\begin{eqnarray}
  M_1&=&M_c(1+\epsilon_1)\nonumber\\
  M_2&=&M_3/1.37\nonumber\\
  m_i&=&M_c(1+\epsilon_i)\nonumber\\
  A_{ijk}&=&-\frac{1}{\sqrt{3}}(m_{i}+m_{j}+m_{k}),\label{softlgvol}
\end{eqnarray}
where $\epsilon_i$ were randomly generated within a domain
$0<\epsilon_i <\epsilon_0$. A reasonable value for $\epsilon_0$
can be taken to be 0.2 as in \cite{Conlon:2007xv}. The low scale
gaugino mass ratios calculated from the above boundary conditions
are $(1.5-2):2:6$. As we see, the ratio of $M_2$ and $M_3$ is
fixed but the ratio of $M_1$ and $M_2$ or $M_3$ is not completely
fixed. This can be seen from (\ref{softlgvol}) as open string
fluxes lead to uncertainties for $M_1$. At high scale, gaugino
masses and squark masses are roughly the same, and are both
boosted by the SU(3) interaction when RG evolved to the low scale.
So generically the gluino is the heaviest particle and the first
and second generation squarks are only a little lighter than the
gluino. The tau slepton is extremely light (close to the mass of
the LSP) which is needed to not overclose the universe by the bino
LSP relics.

\subsubsection{Fluxless $M$ theory $G_2$-MSSM vacua ($\,G_2\,$)}

The M-theory vacua we consider here follow reference
\cite{Acharya:2006ia}-\cite{ABKKS2}. One studies fluxless $M$
theory compactifications on $G_2$ manifolds with at least two
hidden sectors undergoing strong gauge dynamics, at least one of
which has charged matter. This leads to a stabilization of all
moduli and the spontaneous breaking of supersymmetry. The
supersymmetry breaking is dominated by the hidden sector meson
field $\phi$, which is not sequestered from the visible sector.
The gauge kinetic function is a linear combination of all
geometric moduli $s_i$. For the case where the matter K\"{a}hler
metric does not depend on $\phi$, the gaugino masses receive
comparable contributions from moduli and anomaly mediation, but is
different from mirage mediation in the KKLT string-susy model. The
high scale gaugino masses have the following form:
\begin{eqnarray}
M_{a}&\approx&-\frac{1}{4\pi(\alpha_M^{-1}+\delta)}\left\{b_a+\left(\frac{4\pi\,\alpha_M
^{-1}}{\P}-b_a'\phi_0^2\right)\Big(1+\frac{2}{\phi_0^2(Q-P)}\Big)
\right\}\, m_{3/2} \label{gauginohigh} \\ {\rm where}\quad \; b_1&=&33/5,\quad
b_2=1.0, \quad b_3=-3.0, \quad b_1'=-\frac{33}{5}, \quad
b_2'=-5.0, \quad b_3'=-3. \nonumber
\end{eqnarray}
$\alpha_M$ is the tree-level universal gauge coupling and $\delta$
is the threshold correction from the Kaluza-Klein modes. $\phi_0$
is the vev of the meson field, which is of order unity. $P$ and
$Q$ are the ranks of the hidden sector gaugino condensation
groups. Low scale supersymmetry can be obtained only if $Q-P=3$.
To get dS vacua, it is also necessary that the combination
$\P\equiv P\ln(A_1 Q/A_2 P)$ is less than 84, and larger than
about 50 to get a gravitino mass below 100 TeV. The scalar masses
are about equal to $m_{3/2}$ as there is no sequestering in
general. Thus, the soft supersymmetry breaking pattern is such
that there is a large mass splitting between gauginos and scalars,
and the low energy effective theory at the weak scale is mainly
determined by the gaugino sector. Unlike split-SUSY, the higgsinos
in these vacua are as heavy as scalars and also decoupled. This
gives the low scale gaugino masses large finite threshold
corrections from the higgs-higgsino loop. Generically the wino is
the LSP for $G_2$-MSSM models with light spectra, but a wino-bino
mixture is also allowed particularly for heavier spectra.

\subsubsection{Comments on the KKLT and LARGE Volume vacua}
The three classes of type IIB MSSM vacua we described above are
related in some ways. As seen in Sections \ref{Sec:KKLT} and
\ref{Sec:KKLT-2}, KKLT-1 and KKLT-2 are different in the
supersymmetry breaking and uplifting mechanism and they
generically give rise to different soft supersymmetry breaking
terms. KKLT-2 in some sense a broader class of constructions as
the explicit structure of the hidden sector is not completely
specified, while that in KKLT-1 is completely specified. Thus it
is in principle possible to make a model within the KKLT-2 class
which has exactly the same soft supersymmetry breaking terms as
KKLT-1. This happens for example when the hidden section is
sequestered from the visible section and $\zeta_i$ goes to $1/3$
as has been already shown. So, in terms of the soft susy breaking
pattern, models of KKLT-1 are a subset of those of KKLT-2. One
should keep in mind that this overlap is a ``theoretical
overlap"\footnote{in the sense that the two class of models give
rise to the same soft terms from a theoretical point of view.} and
cannot be distinguished at the LHC. So in the distinguishibility
analysis in the rest of the paper, we will focus on the region of
KKLT-2 which does not overlap with KKLT-1. Further study is needed
to learn whether cosmological or additional visible sector physics
can distinguish these constructions phenomenologically.

The KKLT and LARGE Volume MSSM vacua (described in \ref{Sec:KKLT}
and \ref{Sec:LGVOL}) are two distinct regions in the type IIB
landscape. However they can be smoothly connected by dialing some
parameters. The same scalar potential which gives rise to the
large-volume minimum also has a KKLT minimum if $W_0 \ll 1$
\cite{LargeVolume}. As $W_0$ decreases, the two minimums approach
each other and eventually merge. So one could in principle start
with LARGE Volume MSSM vacua and decrease $W_0$ while keeping
$m_{3/2}$ fixed by decreasing the volume $\cal V$. In this way,
the large-volume minimum will gradually lose its ``large volume"
property and become more and more like a KKLT vacua with TeV scale
gravitino mass. We do not know much about the properties of these
intermediate vacua. One possibility is that they lead to a set of
soft supersymmetry breaking terms which interpolate in between the
LARGE Volume MSSM vacua and KKLT MSSM vacua. Even if this is true,
phenomenologically it is not clear whether these intermediate
vacua will survive after all kinds of experimental and consistency
constraints. It may be that a continuous set of intermediate vacua
consistent with data and theory do not exist.

\section{Footprint of ``String-Susy Models" at the LHC}

\subsection{How to construct a Footprint in general}

As was explained in section IIA, a string-susy model is specified
by a set of microscopic parameters characteristic of the class of
string vacua. A complete analysis for the whole (microscopic)
parameter space is necessary if one hopes to discriminate between
different string-susy models. The prediction of a given
string-susy model at the LHC is a map from this parameter space to
LHC signature space. This is a multi-dimensional region which we
call the ``footprint" of the particular string-susy model. In this
paper we construct a footprint for the three string-susy models
described earlier - KKLT MSSM vacua (two variations), LARGE Volume
MSSM vacua and $G_2$-MSSM vacua. We first start with a general
discussion of constructing a footprint of any string-susy model.

As seen in the last section, a set of MSSM soft supersymmetry
breaking parameters can be obtained at the compactification
scale\footnote{This is typically the unification scale}. Below
this scale, heavy stringy and Kaluza-Klein states decouple and the
soft parameters of the MSSM fields are governed by the MSSM
renormalization group equations. Gauge couplings and Yukawa
couplings are determined from current experimental data. The $\mu$
parameter is determined by the correct electroweak symmetry
breaking and $Z$ boson mass. It would of course be preferable to
calculate $\mu$ and $\tan\beta$ from the microscopic theory as
well, but that is not yet possible. In the $G_2$ case $\tan\beta$
is calculated from the theory, but not $\mu$.

In addition the low scale soft supersymmetry spectrum is subject
to various constraints from current observation. There are lower
bounds on the masses of the various sparticles from the SUSY
searches at LEP and Tevatron. The most important ones are the
chargino mass limit and the higgs mass limit. There are also
constraints from observations in cosmology, i.e. dark matter relic
abundance $\Omega h^2$. Although one can compute the thermal relic
density reliably, there may be other contributions from
non-thermal production or other unknown sources, so one should
impose an upper limit constraint but not a lower limit one.

In order to connect to LHC experiments, the next step is to
simulate the $p\,p$ collision and the decay of particles produced
at the LHC followed by detection of the surviving particles in the
final state. In our analysis, we use the PGS4 \cite{pgs} package
which generates events using PYTHIA6.4 \cite{pythia} and then
perform the detector simulation, where the default configuration
of the detector parameters are used. While PGS is not a fully
realistic detector simulator for the LHC, it is simple, fast and
gives a ``pretty good" simulation result, as its name suggests.
The result from PGS usually agrees fairly well with the result one
might obtain with a full-fledged detector simulation. In many
cases the agreement is good, of the order of 20$\%$. To construct
the footprint of the string-susy models described earlier, we
sample the high-scale (microscopic) parameter space with a large
number of points. For collider phenomenology, the simplest
assumption of equal probability distribution on the parameter
space is presumably sufficient. For each point we sample, the
corresponding signatures are computed through the aforementioned
procedure. For our purposes here, where we compare predictions of
different models, these procedures are adequate. Later the
analysis can be sharpened. The procedure to go from string/$M$
theory to signatures may seem complicated, but now user-friendly
softwares exist to do that. One can access much of the software
through the LHC Olympics website \cite{LHCO}.

To demonstrate the general approach shown above, we construct
footprints of the four string-susy models discussed for an
integrated luminosity of 5 $fb^{-1}$. In the simulation, we use
the L2 trigger in PGS to get better S/B ratios \cite{LHCO}. For
signatures, we use the following selection cuts for objects in
each event
\begin{itemize}
\item{Jet $P_T > 50$ GeV; Lepton and Photon $P_T
> 10$ GeV; $\displaystyle {\not} E_T > 100$ GeV}.
\end{itemize}
This means only objects satisfying the above cuts are kept in the
event record. For backgrounds, we use the background sample in the
LHC Olympics webpage \cite{LHCO} which includes dijets, $t\bar t$
and $W,Z$+jets processes and scale it up to get an estimate of the
background for an integrated luminosity of 5 $fb^{-1}$. Other
backgrounds may be important and should be taken into account in a
more thorough analysis of the backgrounds, but as will be seen
below the treatment of backgrounds will not have much affect on
our main results. The condition for a (counting) signature to be
observable above the background is:
\begin{eqnarray}
  \frac{S}{\sqrt{B}} > 4, \quad S > 5,
\end{eqnarray}
where $S$ is the number of signal events that pass the selection
cuts, while $B$ is number of background events that pass the same
cuts. Thus we can assign an observable limit for each (counting)
signature below which the signal is not likely to be observed.

\begin{figure}[h!]
  \begin{eqnarray}
  \begin{array}{c}
  \includegraphics[width=380pt]{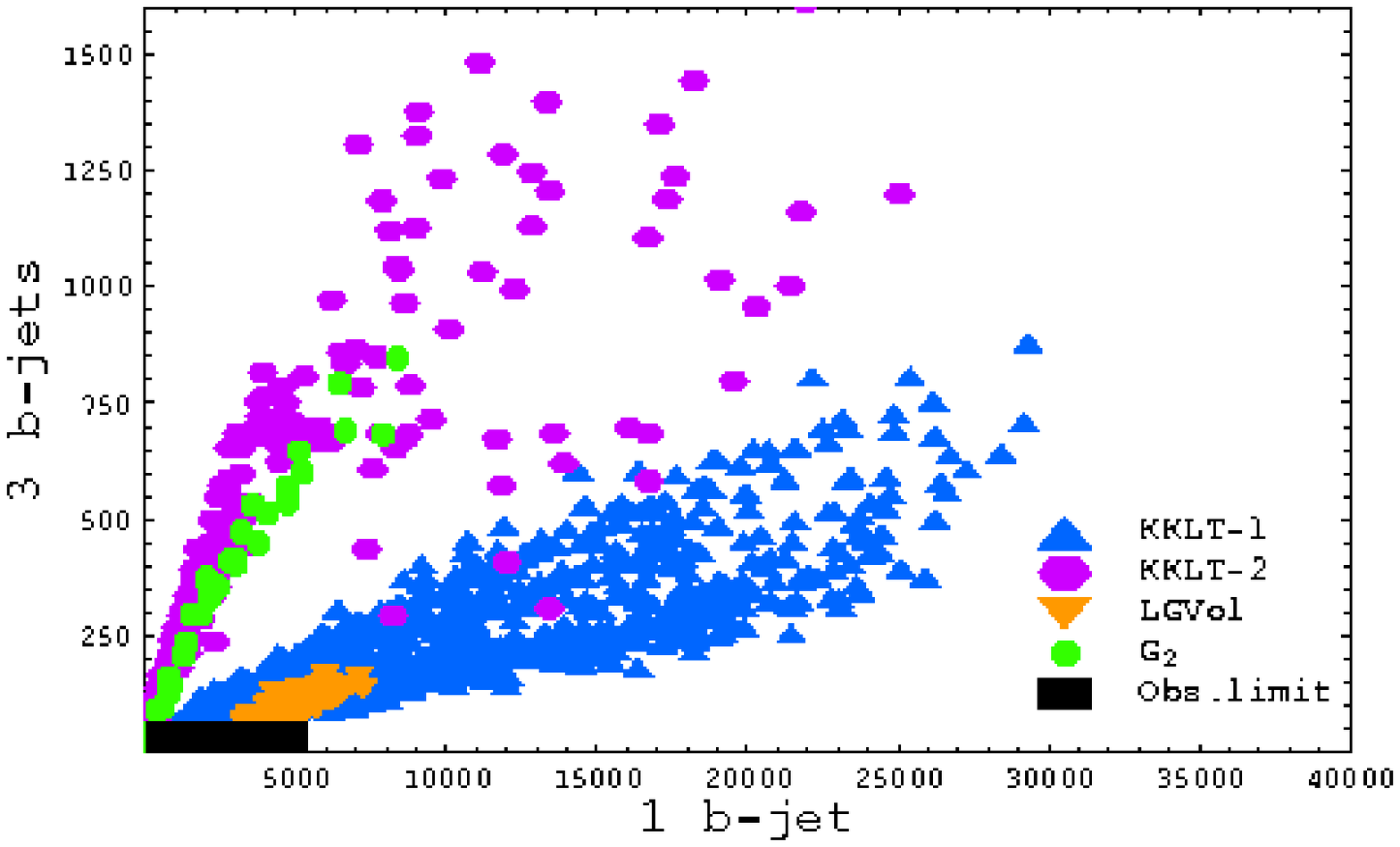}\\
  \includegraphics[width=380pt]{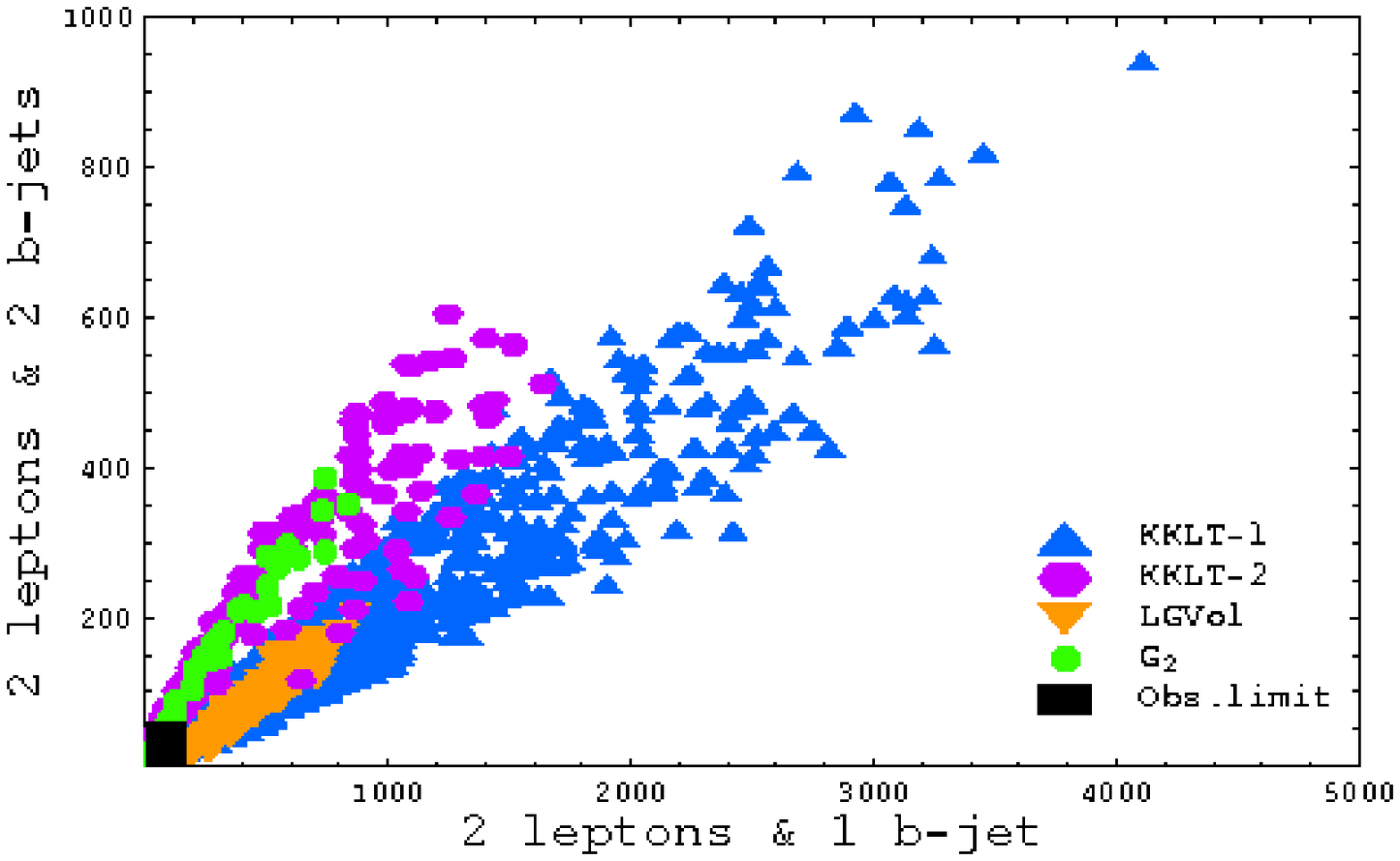}
  \end{array}\nonumber
  \end{eqnarray}
  \caption{Two-dimensional slices of the footprint of the three
  string-susy models. All models are simulated with 5$fb^{-1}$ luminosity in PGS4 with L2 trigger.
  If not explicitly stated, all signatures include a least two hard jets and large missing transverse energy.
  For each example, the points are generated by varying the microscopic parameters over their full ranges,
  as explained in Section \ref{Sec:scan}.}
  \label{fig:plots1-1}
\end{figure}

\begin{figure}[h!]
  \begin{eqnarray}
  \begin{array}{c}
  \includegraphics[width=380pt]{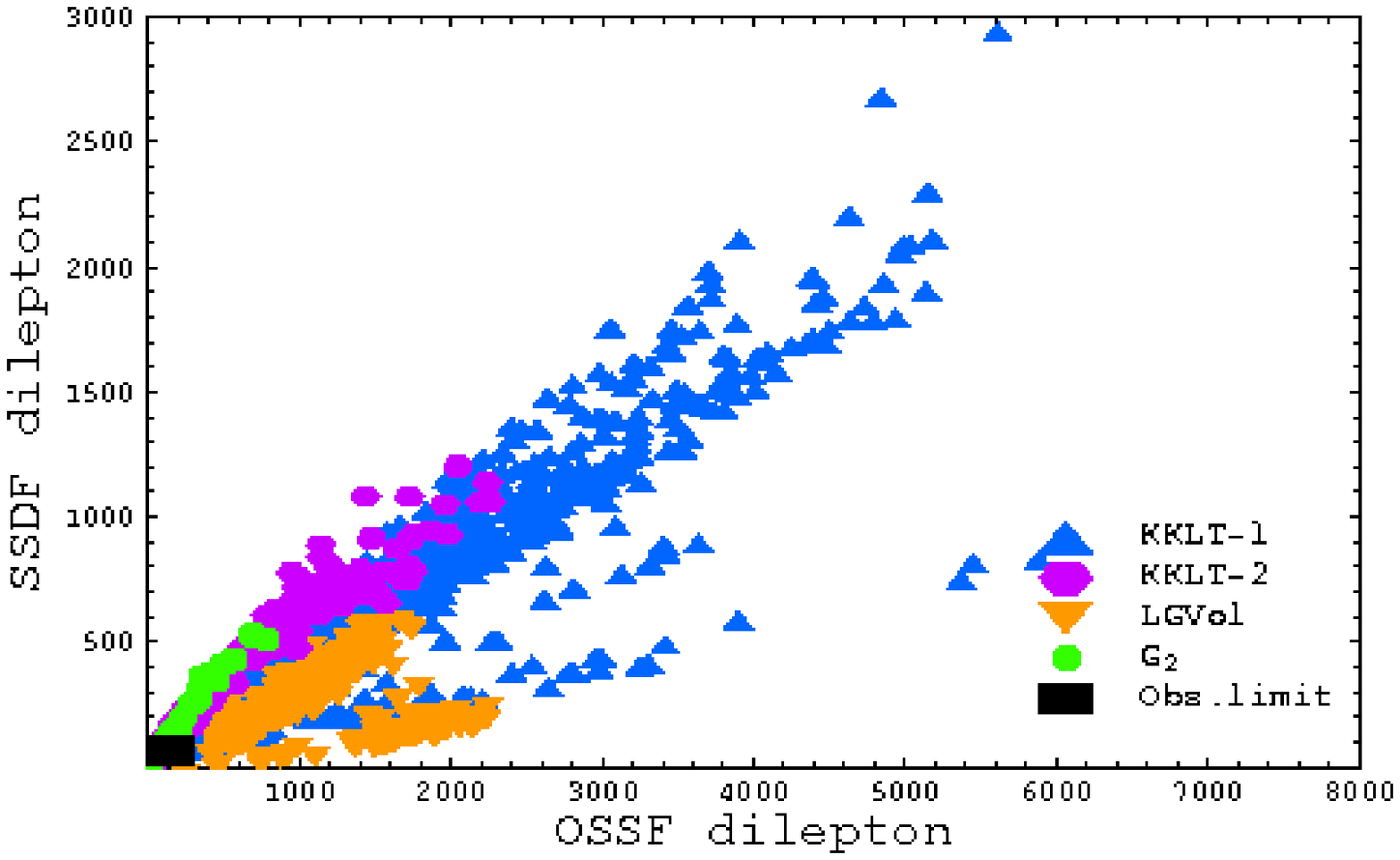}\\
  \includegraphics[width=380pt]{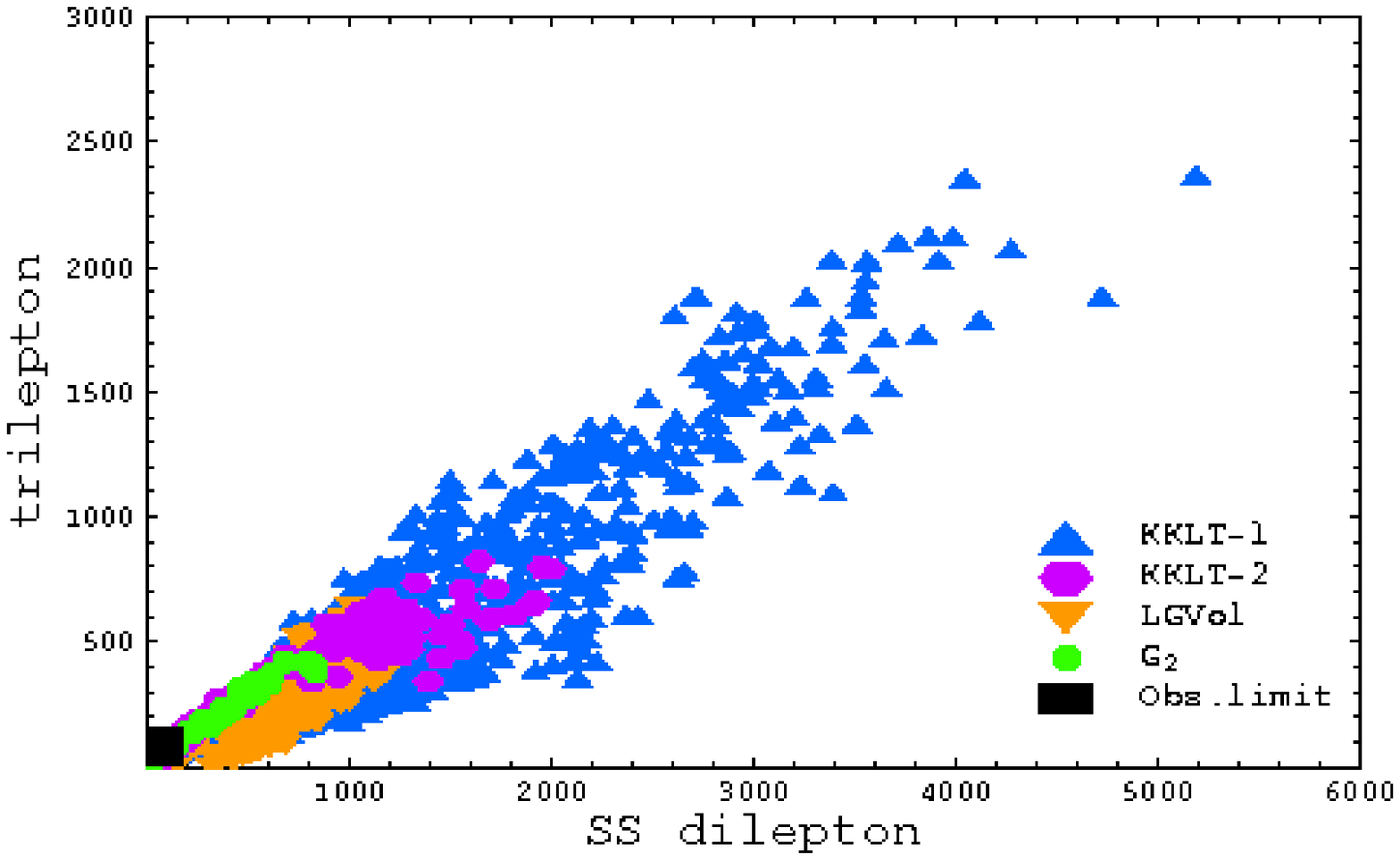}
  \end{array}\nonumber
  \end{eqnarray}
  \caption{Two-dimensional slices of the footprint of the three
  string-susy models. All models are simulated with 5$fb^{-1}$ luminosity in PGS4 with L2 trigger.
  If not explicitly stated, all signatures include a least two hard jets and large missing transverse energy.
  For each example, the points are generated by varying the microscopic parameters over their full ranges,
  as explained in Section \ref{Sec:scan}.}
  \label{fig:plots1-2}
\end{figure}

Figures \ref{fig:plots1-1} and \ref{fig:plots1-2} represent some
simple 2D slices of footprints. There is no particular reason for
the above choice of signature plots, they are just meant to
illustrate general features. Some of these plots however have the
added advantage that they are also useful in distinguishing some
string-susy models. In all the (counting) signature plots, the
approximate regions where the SM dominates are entirely blacked
out \footnote{More precisely, it is the region bounded by the
observable limits}. Immediately one can see from these plots that
the footprints for these string-susy models are finite regions in
signature space. This implies that a well-defined string-susy
model is not likely to cover the whole signature space, but only a
part of it. In addition, based on these footprints, one can
readily distinguish among the string-susy models in many cases.
For example, the plot of 1-b jet vs. 3-b jets clearly separates
the KKLT-1 and $G_2$ string-susy MSSMs. By definition a footprint
covers all possible signatures that might come out from a
string-susy model. Therefore plots of footprints demonstrate the
overall difference between different string-susy models. In this
sense the footprint analysis generalizes the familiar benchmark
analysis, where one does not get an overall picture of signatures
for a given model thereby making it difficult to distinguish two
classes of models. We emphasize \emph{that a (n-dim) footprint is
the full region on a (n-dim) signature plot or any 2-D slice
generated as the microscopic parameters are varied over their
entire allowed ranges respectively.}

\subsection{Generic Features of Footprints}

Some generic features of these footprints can be easily understood
as follows. For simplicity, we will only focus on simple counting
signatures, which illustrate many of the important points we want
to emphasize. Counting signatures are always bounded by the
maximum cross section, which is related to existing lower limits
on masses. Hence the 2D projection of a footprint for counting
signatures must be bounded along the radial direction. In
addition, if no upper bound is imposed on sparticle masses, the
footprint can continuously approach the origin. However the region
below the observable limit is not interesting.

The angular dispersion of the footprint is due to the variation in
the spectrum, which leads to a variation in branching ratios and
in turn, the signatures. The smaller the angular dispersion the
larger the correlation in the low scale soft spectrum and the more
predictive the string-susy model. However, the exact spread
depends on the particular signatures used because of many factors.
For example, even a completely random MSSM soft spectrum will not
cover the entire angular range from $0$ to $\pi/2$. One also has
to take into account real-world ``detector effects". For example,
even for a model with no b-quark produced at parton level, there
could be some b-jets in the final data since other quarks could be
mistagged as b-quarks. For PGS4 loose b-tag, the charm quarks are
mistagged as b-quarks with a probability 13\%, while for other
quarks the probability is about 1\%. This implies that the ratio
of b-jets to jets is at least 1\%. In cases with significant charm
quark production, the fake b-jets will be even larger. Another
example is that a $k$-quark final state at parton level can be
read-off as an event with $k-1$ jets if one of the jets is soft
and thereby fails to pass selection cuts, or if two jets merge
together to form a single jet. The limited statistics of our
simulations are not a problem since the precise regions are not
needed for our main conclusions, and since we understand why the
regions have the boundaries they do. Later, analysis with more
statistics can be done.

\subsection{Origin of Distinguishibility - Correlations}
It is desirable and important to qualitatively understand the
footprint boundaries and the difference between footprints. The
features in footprints can be connected to the underlying theory
by understanding correlations between soft parameters which in
turn have their origin in the structure of the underlying theory.
One could understand this as follows. Formally, the collider
signatures ($s_i$) are functions of the MSSM masses and couplings
(call them $m_i$ in general), which are themselves parameterized
by the underlying ``microscopic'' parameters (call them $\xi_k$).
One has:
\begin{eqnarray}
  s_i=s_i(m_j)=s_i\Big(m_j(\xi_k)\Big).
\end{eqnarray}
For an arbitrary set of MSSM parameters, one would get a very
broad set of signatures\footnote{The signatures will still not be
uncorrelated due to the structure of the MSSM itself and also due
to detector effects. However, we are interested in correlations
which are present in addition to these.}, or equivalently, the
corresponding footprint would cover a very large region in
signatures space. However if there is a non-trivial dependence on
the more fundamental (microscopic) parameters $\xi_k$, the MSSM
parameters are correlated with each other and so are the
signatures. Therefore by understanding how correlations between
soft parameters are connected to the structure of the underlying
theory, one can understand why a given footprint occupies a given
region in signature space and not some other region. In order to
make the task easier, it is helpful to first understand the
footprint in terms of the pattern of spectra of the class of
models and then understand the spectra in terms of the soft
parameters determined from the underlying theory. For simplicity,
here we only explain the former. The latter can also be done in a
straightforward manner, the interested reader can refer to the
references available for the string-susy models studied here. The
features described here are not used in constructing the plots,
but are valuable in understanding the plots.
\begin{figure}[h!]
  \begin{eqnarray}
  \begin{array}{c}
  \includegraphics[width=400pt]{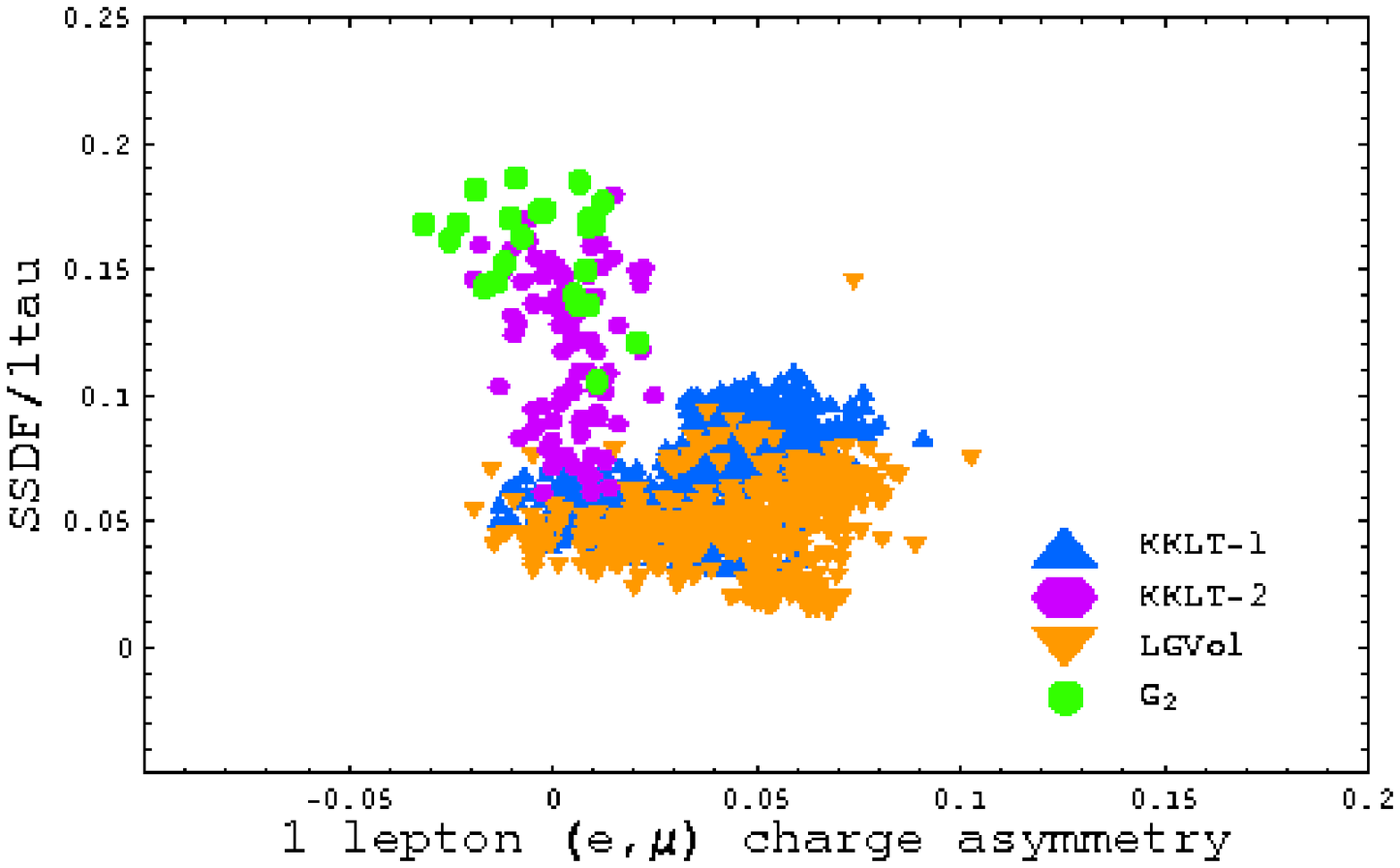}
  \end{array}\nonumber
  \end{eqnarray}
  \caption{A particular slice of footprint for the models studied.
  The one-lepton charge asymmetry (only include $e$ and $\mu$) is defined as
  $A_c^{(1)}\equiv\frac{N_l^{+}-N_l^-}{N_l^{+}+N_l^-}$. The
  SSDF/1tau signature\protect\footnote{Ratios of counting signatures are
  independent of the total rate and are sometimes useful. This gives an example.}
  is defined as the ratio of the number of events with SSDF
  dilepton and the number of events with 1 tau lepton. All models are simulated
  with 5$fb^{-1}$ luminosity in PGS4 with L2 trigger.
  If not explicitly stated, all signatures include a least two hard jets and large missing transverse energy.
  For each example, the points are generated by varying the microscopic parameters over their full ranges,
  as explained in Section \ref{Sec:scan}.}
  \label{asymm}
\end{figure}

Let us start with colored particle production. For KKLT-1 and
LARGE Volume string-susy models, the squarks are lighter than the
gluino and the dominant production is squark pair production and
the squark-gluino production. For $G_2$-MSSM models, the dominant
production is gluino pair production since all squarks are
extremely heavy. The KKLT-2 models also have a large scalar mass
and are dominated by gluino pair production. This difference in
the dominant production channel already leads to a difference in
the lepton-charge asymmetry, as seen in Figure \ref{asymm}. Below
we list some broad distinguishing features in the spectra of the
string-susy models and their related signatures.
\begin{enumerate}\label{enu:features}
    \item Sleptons in KKLT-1 and LARGE Volume MSSM models are relatively light
    (lighter than the gluino). Moreover $\tilde \tau$ is generically the
    lightest slepton. On the other hand, sleptons in $G_2$-MSSM models are very heavy
    (around ${\cal O}(10)\TeV$).
    So, signature plots sensitive to lepton flavor asymmetry could differentiate KKLT-1 and LARGE Volume from $G_2$.

    \item The gaugino mass ratios are different for different models, which lead to a difference in
    the jet multiplicity. For KKLT-1, the difference between $M_3$ and the LSP mass is much
    smaller than that of LARGE Volume and $G_2$ models (for the same gluino mass). So if
    we use a hard $p_T$ cut on the jets (e.g. $P_{T}(jet)\ge 200\GeV$), then most of the four-jet events
    in KKLT-1 cannot pass the cuts since they are mostly from gluino pair production.
    However, for two-jet events, since the mass difference $m_{\tilde q}-M_{LSP}$ is
    large enough, most of them will pass the cuts. Thus we can probably use signature plots
    of events with 2 jets and events with 4 jets to distinguish models of KKLT-1 with those of LARGE
    Volume and $G_2$. However in our plots, it can be seen there is still
    an overlap between KKLT-1 and LARGE Volume regions. The KKLT-1 models
    in the overlap region are exactly those with heavier gluinos.

    \item Since the top Yukawa coupling is large, from RGE running the
    stop is lighter compared to other squarks. This is more pronounced for $G_2$ models
    since the $\tan\beta$ is particularly small ($\sim 1.5$).
    The gluino will preferentially decay via a virtual stop and lead to
    b-rich events. For KKLT-1 and LARGE Volume models, the stop is
    again light and its production rate is big. However
    since all other squarks are also copiously produced, the
    overall branching ratio for the events with b-jets is not
    particularly large, so signature plots involving numbers of b-jets can
    distinguish $G_2$ models from those of KKLT-1 and LARGE Volume.
    In cases with very light stops, the stop production rate is very large,
    but they will decay into charm quarks instead of bottom quarks if the stop
    is lighter than ${\tilde C}_1$ \footnote{In this case, the
    decays
    $\tilde t\rightarrow b {\tilde C}_1$ and $\tilde t\rightarrow t {\tilde N}_1$ are kinematically closed. }.
    Therefore these models will again give relatively small number of events with b-jets.

    \item The KKLT-2 models appear in the signature space similar to
    $G_2$ models because they both have very heavy scalar masses.
    However they extend to a much bigger region in many plots and have
    some overlap with KKLT-1 models. Because of the large scalar
    masses, KKLT-2 models can be in the focus point region which is
    consistent with the dark matter constraints. This means in these
    models the LSP has a significant higgsino component. We also know
    that because of the higgsino mixture, gluinos tend to decay into
    b-jets through the large Yukawa couplings. As the consequence of
    the variation in the higgsino fraction, there is a large spreading
    in the b-jet signatures.
\end{enumerate}

\begin{figure}[tb]
  \begin{eqnarray}
  \begin{array}{c}
  \includegraphics[width=390pt]{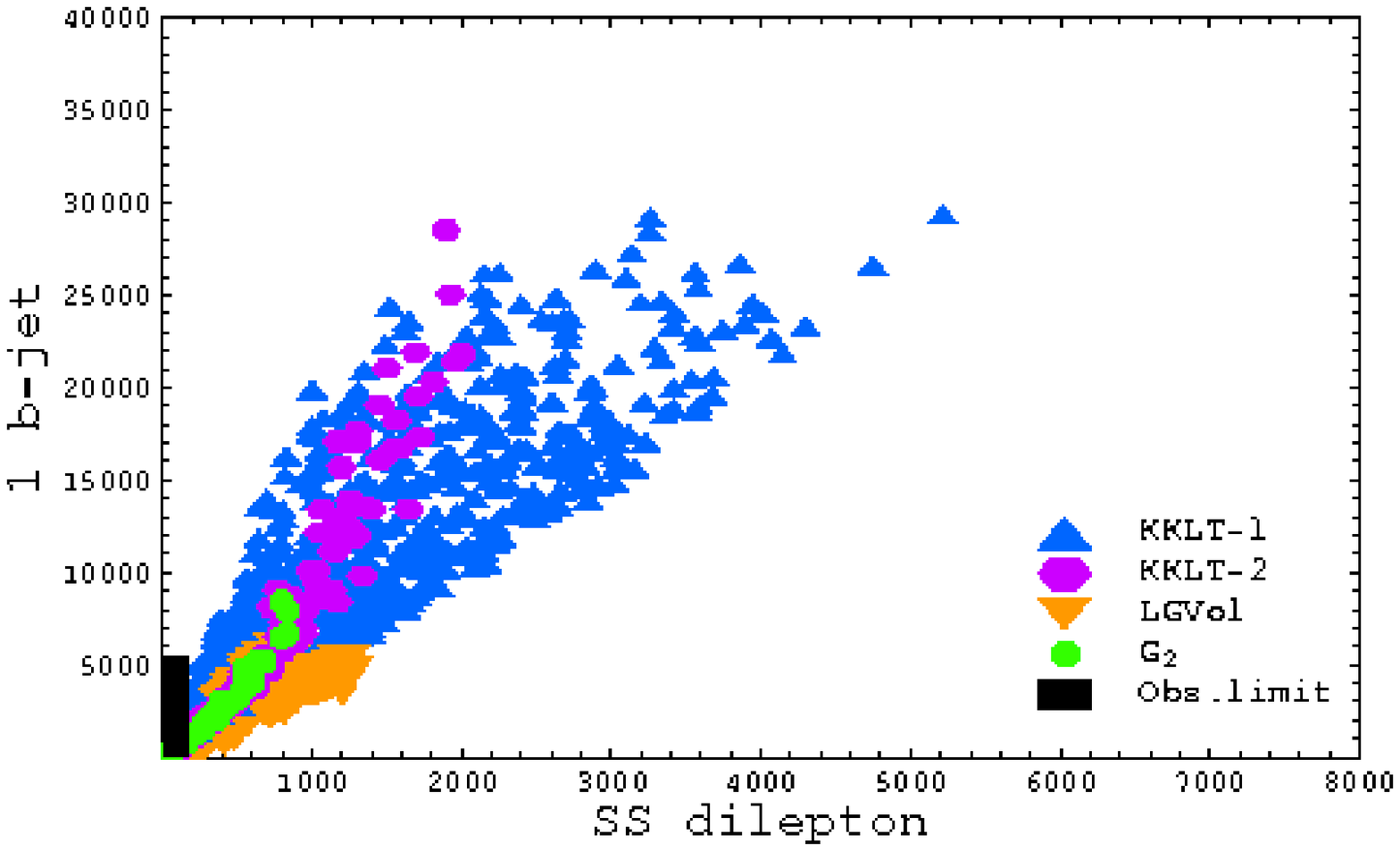}\\
  \includegraphics[width=390pt]{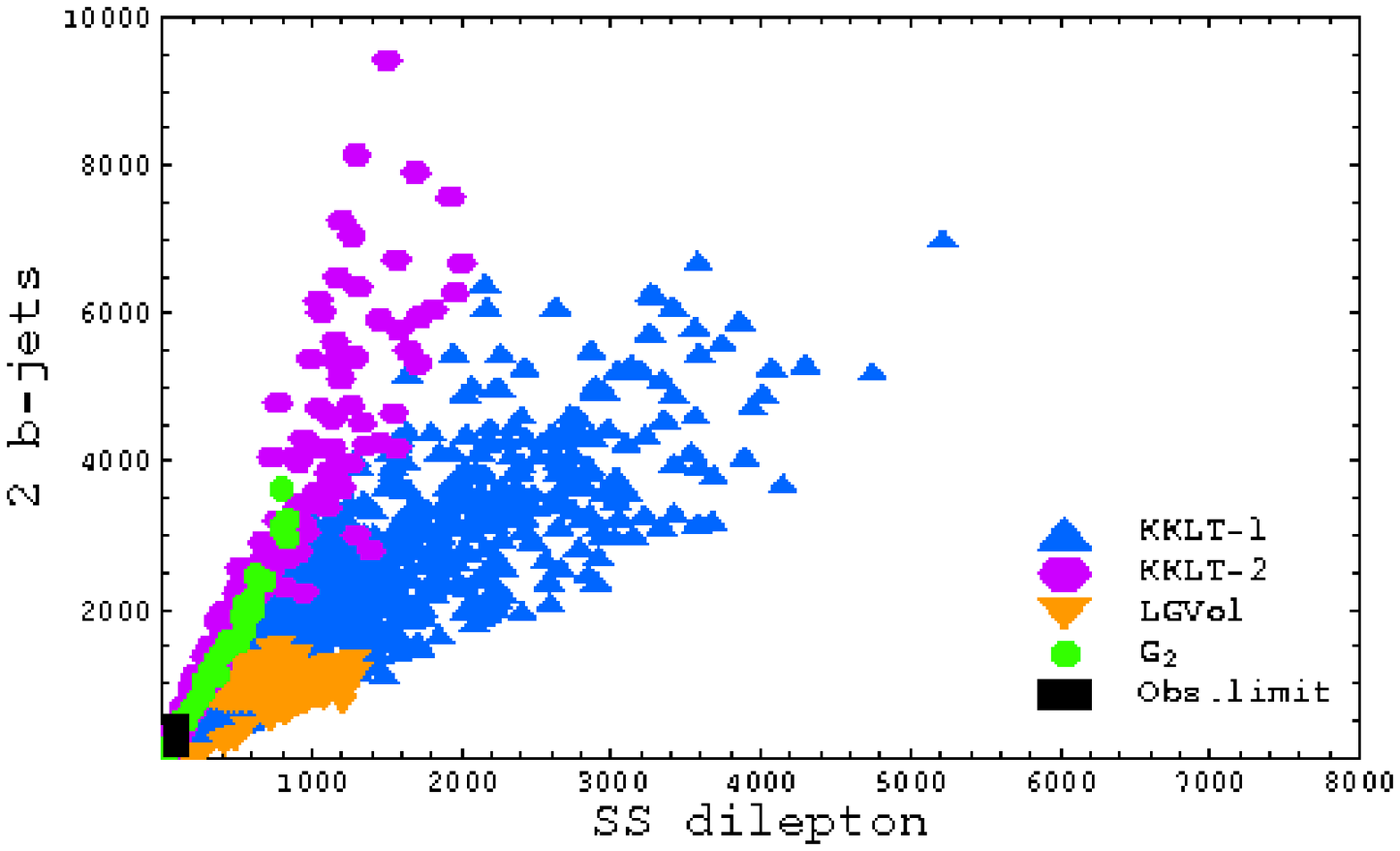}
  \end{array}\nonumber
  \end{eqnarray}
  \caption{Slices of footprints for the models studied.
  All models are simulated with 5$fb^{-1}$ luminosity in PGS4 with L2 trigger.
  If not explicitly stated, all signatures include a least two hard jets and large missing transverse energy.
  For each example, the points are generated by varying the microscopic parameters over their full ranges,
  as explained in Section \ref{Sec:scan}.}
  \label{fig:plots1-3}
\end{figure}

\begin{figure}[ht]
  \begin{eqnarray}
  \begin{array}{c}
  \includegraphics[width=390pt]{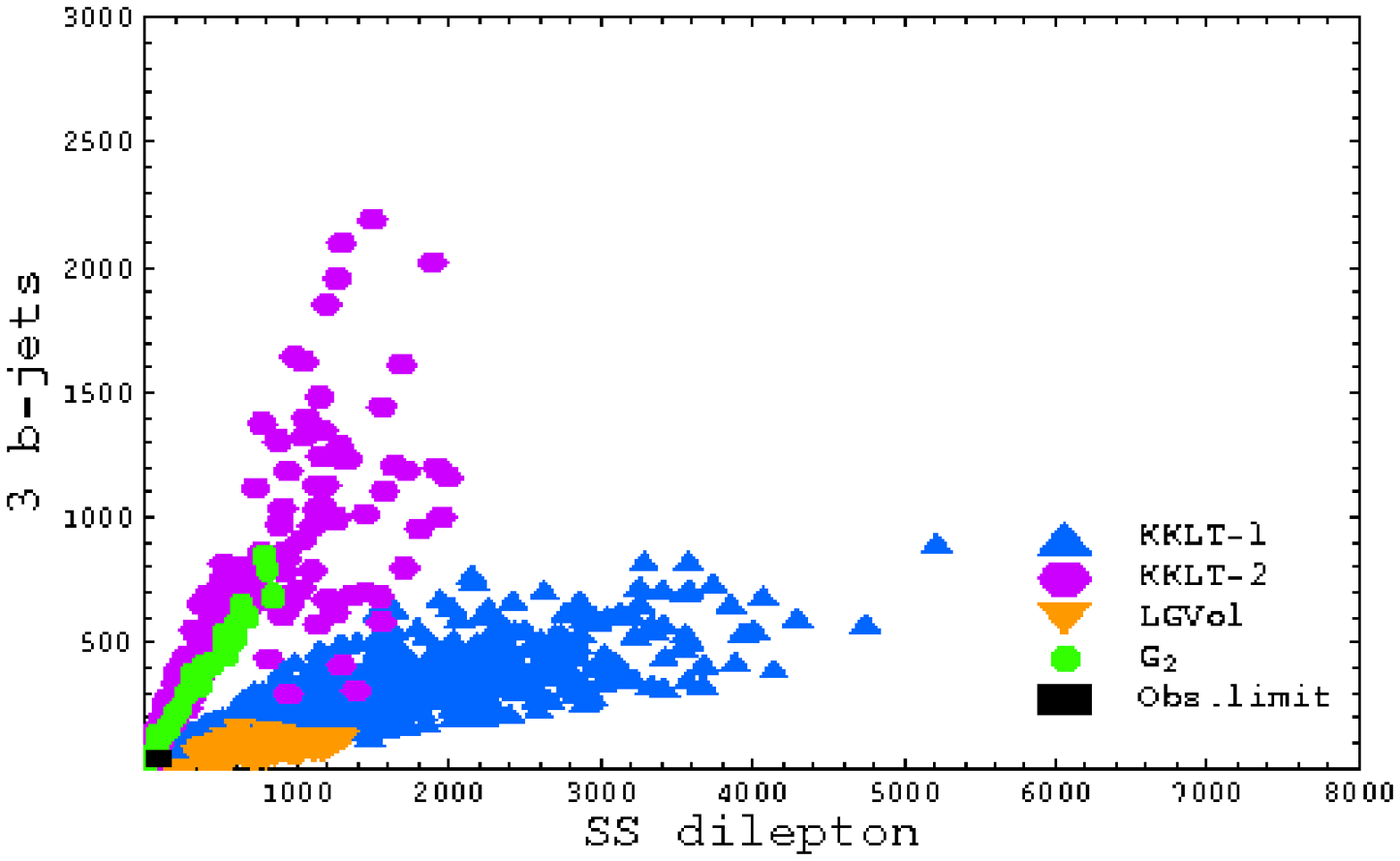}\\
  \includegraphics[width=390pt]{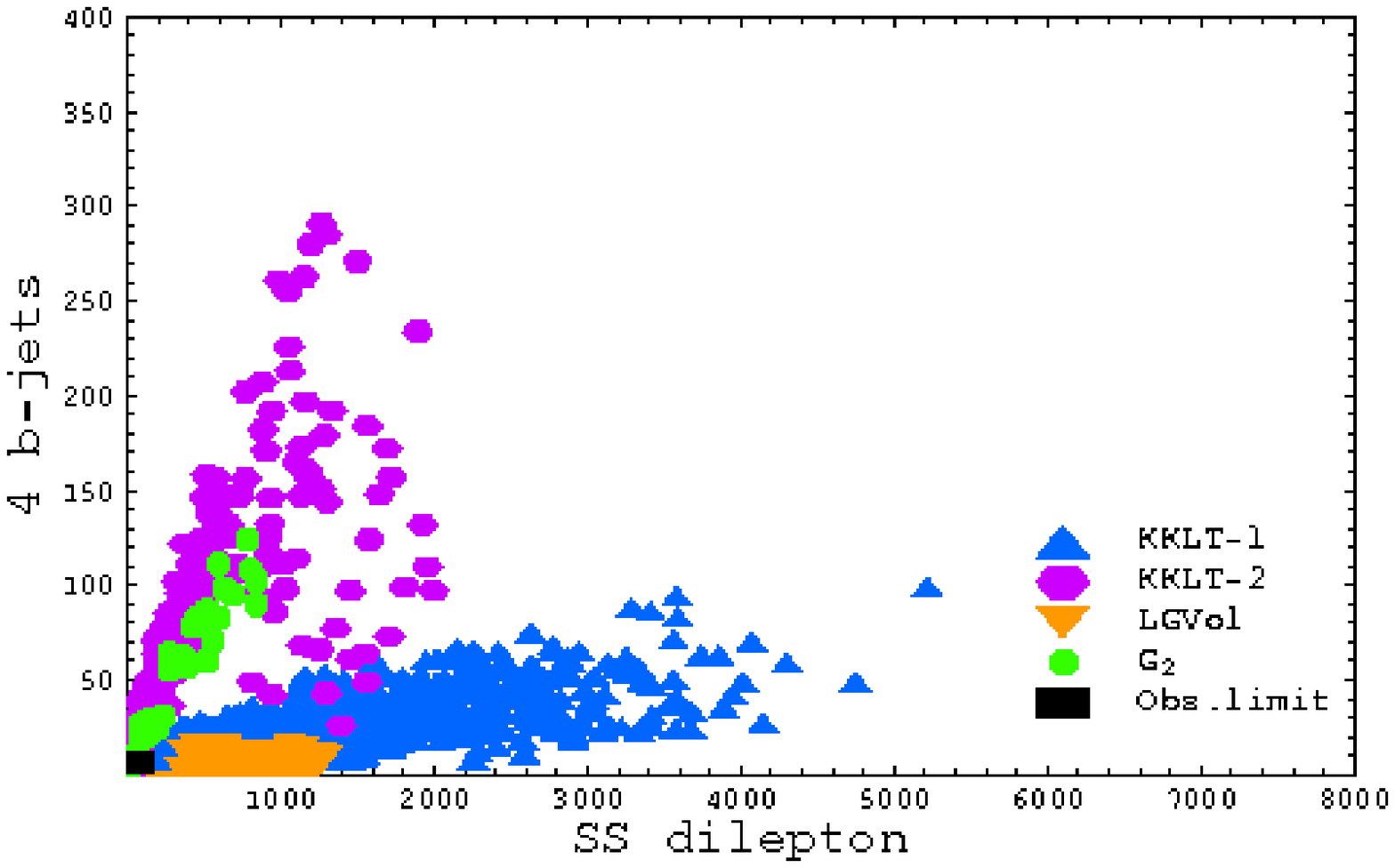}
  \end{array}\nonumber
  \end{eqnarray}
  \caption{Slices of footprints for the models studied.
  All models are simulated with 5$fb^{-1}$ luminosity in PGS4 with L2 trigger.
  If not explicitly stated, all signatures include a least two hard jets and large missing transverse energy.
  For each example, the points are generated by varying the microscopic parameters over their full ranges,
  as explained in Section \ref{Sec:scan}.}
  \label{fig:plots1-4}
\end{figure}

In Fig. \ref{fig:plots1-3} and \ref{fig:plots1-4}, we show how the
b-jet multiplicity affects the relative positions of these
footprints. It can be seen that as the b-jet multiplicity
increases, the footprints of $G_2$ and KKLT-2 models become
isolated from the other two with a larger angular separation. This
demonstrates that the b-jet multiplicity is related to a certain
structure of the underlying theory. Thus, this is an example of a
signature which is directly correlated with underlying structure
of the theory and is particularly useful. For KKLT and LARGE
Volume models, their relative position does not change much as the
b-jet multiplicity changes, implying they have similar structure
with regard to b-jet multiplicity. To summarize, the boundaries
and the distinguishibility of footprints can be understood in
terms of the spectra, and in turn in terms of the correlations
between the soft supersymmetry breaking parameters determined by
the underlying string-susy model.

Finally one should remember that although in principle we can have
a large number of signatures, they are not orthogonal to each
other. It was shown explicitly in \cite{ArkaniHamed:2005px} that
for a set of MSSM models with 15 parameters, the effective
dimensionality of signature space (with 1808 signatures) is only
$\sim$ 5 or 6.
This means differences in the spectra may be lost in the process
of mapping to signatures. This can also be seen in the various
signature plots we have made, where only a few of them are quite
different. The origin of this is related to the nature of hadron
collision where signatures are usually polluted by large
combinatorial backgrounds. Therefore figuring out analytically how
to pick out mutually independent signatures which can distinguish
classes of models is very difficult in general.
In practice, as demonstrated in Section IV, distinguishing classes
of models can be done by adding lots of signature plots since the
overlap region always decreases. By doing this one can identify
useful independent signatures. Sometimes it also helps to pick
signature plots sensibly based on the qualitative features
described above. We will discuss more about this issue in Section
IV A.

\begin{figure}[h!]
  \begin{eqnarray}
  \begin{array}{c}
  \includegraphics[width=390pt]{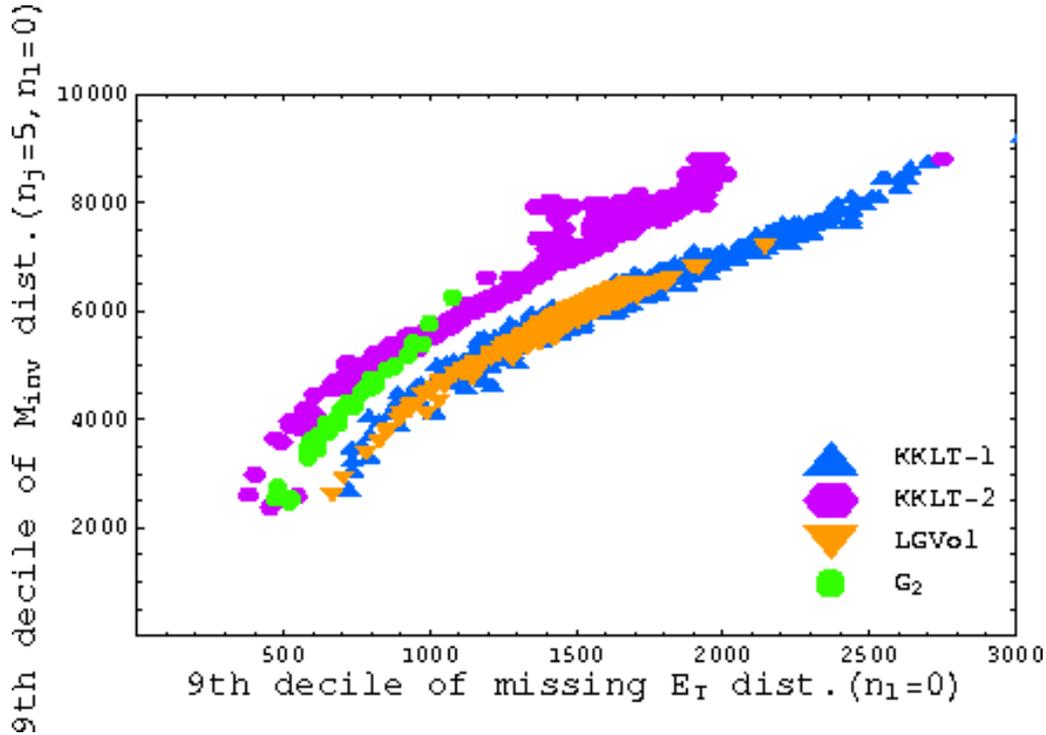}
  \end{array}\nonumber
  \end{eqnarray}
  \caption{Slices of footprints for the models studied.
  All models are simulated with 5$fb^{-1}$ luminosity in PGS4 with L2 trigger.
  For each example, the points are generated by varying the microscopic parameters over their full ranges,
  as explained in Section \ref{Sec:scan}.}
  \label{fig:quantile1}
\end{figure}

We have mostly focused on counting signatures, but one can also
study various distributions. Distributions can be used similarly
to counting signatures if they are converted into quantiles. In
our initial analysis, we have implemented the following basic
distributions:
\begin{itemize}
\item{Effective mass of all objects $m_{eff}=\sum_a P_T^a$,
divided into 12 categories labelled by number of jets and leptons
in the event: $n_j=2,3,4,5^+$, $n_l=0,1,2^+$} \item{Missing $E_T$
distributions, divided into 3 categories labelled by the number of
leptons $n_l=0,1,2^+$}\item{Invariant mass of all objects
$m_{inv}=\left(\sum_a P^a_{\mu}\right)^2$, divided into 12
categories labelled by number of jets and leptons in the event:
$n_j=2,3,4,5^+$, $n_l=0,1,2^+$}
\end{itemize}
\begin{figure}[h!]
  \begin{eqnarray}
  \begin{array}{c}
  \includegraphics[width=390pt]{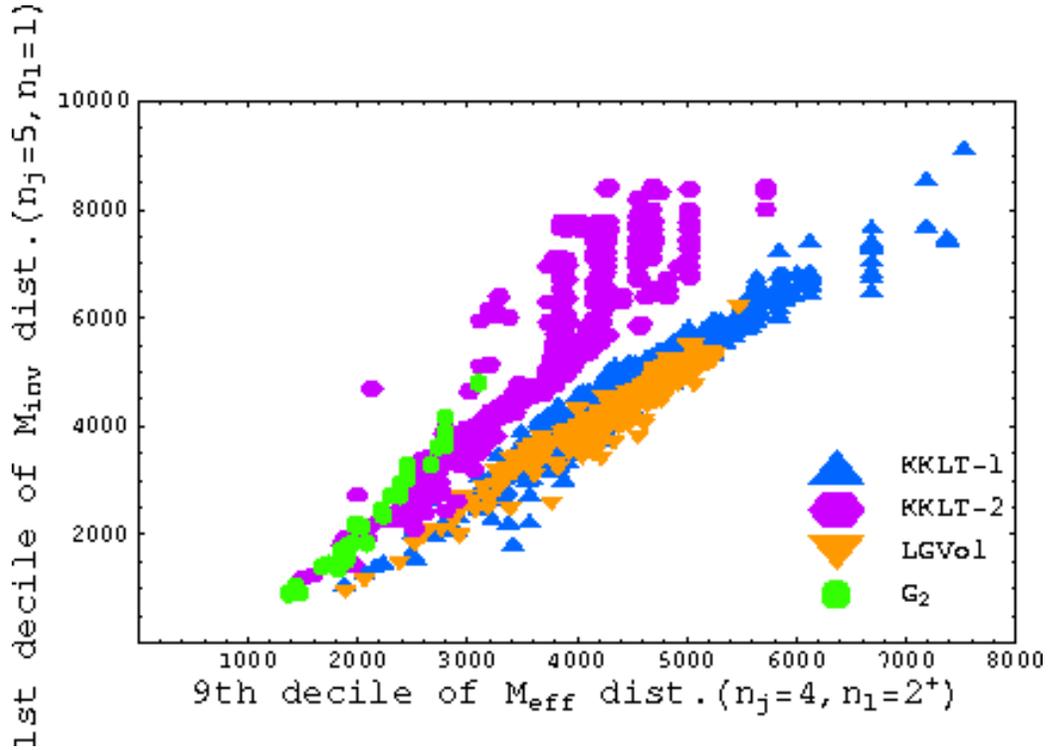}
  \end{array}\nonumber
  \end{eqnarray}
  \caption{Slices of footprints for the models studied.
  All models are simulated with 5$fb^{-1}$ luminosity in PGS4 with L2 trigger.
  For each example, the points are generated by varying the microscopic parameters over their full ranges,
  as explained in Section \ref{Sec:scan}.}
  \label{fig:quantile2}
\end{figure}

To get quantile (decile for example) signatures, the entries in a
distribution are sorted into ten bins such that each bin contains
10\% of the total events. The boundaries of the bins are taken as
signatures, with no signatures related to the lower and upper
boundaries of the whole distribution. Therefore each distribution
gives 9 signatures and we have 243 signatures for the 27
distributions above. We have examined many different quantile
signature plots. There are a few plots which can separate some
string-susy models, two of them are shown in Figures
\ref{fig:quantile1} and \ref{fig:quantile2}.

We have seen not only that footprints are generally limited, but
that we can always get a qualitative understanding of the
boundaries of the footprint region. Therefore one can obtain a lot
of insight even without a high statistics simulation of the
footprint region in many cases, although in some cases when there
is data increasing statistics could turn out to be important.

\subsection{Details of Constructing Footprints for String-Susy
Models}\label{Sec:scan}

In this subsection, we explain in more detail the construction of
the footprint of the various string-susy MSSMs. Starting with the
KKLT-1 string-susy model, the high scale parameter space is
defined in section 2. Explicitly, the low scale soft spectrum is
specified by the following parameters:
\begin{eqnarray}
  M_s, \quad \alpha, \quad n_l, \quad n_q, \quad n_h, \quad \tan\beta, \quad {\rm
  sgn}(\mu).
\end{eqnarray}
To sample the parameter space, one has to set the ranges for
scanning these parameters. First, based on the previous works
\cite{Choi:2005uz,Falkowski:2005ck,Choi:2005hd,Baer:2006id} we
already know that for $\alpha$ below $4-5$, the model is excluded
either by the presence of tachyons or a stop/stau LSP, while for
$\alpha$ above $\sim 10$ the model (satisfying the dark matter
constraint) usually gives rise to a very light spectrum. So we
choose to vary $\alpha$ from $4$ to $10$, which we think covers
the most interesting region of KKLT-1. As can be seen in
Eq.(\ref{KKLTsoft}), $M_s \alpha$ controls the scale of the
sparticle masses. To avoid those cases with very small masses
which are excluded by the SUSY search limits as well as those with
very large masses which are too heavy to be interesting at low
luminosities, we take $M_s$ to be in the range $25\GeV$ to
$100\GeV$. As usual $\tan\beta$ is taken to be from $1$ to $50$
(this is also used for all other models). As mentioned in Sec.
\ref{Sec:KKLT}, modular weights $n_l$, $n_q$ and $n_h$ are not
fixed unless the string construction is explicitly specified. So
in order to be general we allow a continuous variation of the
modular weights from $0$ to $1/2$ independently. In addition, for
a practical step-by-step analysis of the scan of these modular
weights, we divide the task of the complete scan into several
pieces
\begin{eqnarray}
\left\{\begin{array}{ll}
    \textrm {choice 1}: & 0 < n_h, n_q, n_l < 0.1 \\
    \textrm {choice 2}: & 0 < n_h < 0.1, \;\; 0 < n_l, n_q < 0.5 \\
    \textrm {choice 3}: & 0 < n_h, n_q, n_l < 0.5 \\
\end{array}\right.\label{choices}
\end{eqnarray}
The first choice corresponds to a perturbation of
zero-modular-weight models with an uncertainty $0.1$. The size of
the error is somewhat arbitrary but will not affect any of our
conclusions. The second choice is to allow for a large variation
for quark and lepton modular weights. The third one should capture
most of the cases with non-zero-modular weights, as those with
modular weights beyond $0.5$ are very likely to be excluded by the
presence of tachyonic scalars. Clearly the first choice is a
subset of the second one, which is in turn a subset of the third
choice. Each choice is randomly sampled with 500 points.

For a given set of these parameters, we compute the corresponding
soft terms at the unification scale\footnote{we assume that these
models give rise to gauge coupling unification as is suggested
from the unification of couplings in the MSSM.} and then evolve
them using SOFTSUSY 2.0 \cite{softsusy} to the TeV scale to get
the sparticle spectrum. The relic density of neutralino dark
matter is calculated using MicrOMEGAs v1.3.6 \cite{micromegas}.
Only models with $\Omega_{LSP}h^2 < 0.12$ are allowed by the relic
density constraint. In the scan, models with tachyonic scalar
masses as well as those which violate the chargino mass constraint
($m_{\tilde \chi_1} \ge 104\GeV$) or higgs mass constraint ($m_h
\ge 114\GeV$) are rejected. We also impose a cutoff for stop mass
less than $300\GeV$. This is a convenient choice for simulation
due to limited computing time and can be relaxed. It is clear this
cutoff will remove the large cross section region of the
footprint. However it will not affect the essential feature of the
correlations between signatures and therefore, is not very
important for our analysis.

For the KKLT-2 string-susy model, the scalar masses are
generically heavy. To focus on this region, we take parameters
$\zeta_i$ in the range from $0$ to $1/6$. $\zeta_i$ are assumed to
be different for the lepton, quark and higgs fields but the same
among different generations. The parameter $\alpha$ in this model
can have a larger variation, which is taken from $4$ to $20$. The
gravitino mass is chosen to vary from $1\TeV$ to $10\TeV$.

For the footprint of LARGE Volume string-susy models, we generate
sample points by varying $M_3$ from $400\GeV$ to $500\GeV$ and
$\epsilon_i$ from $0$ to $0.2$. The lower value of $M_3$ is chosen
so that the light higgs mass is above the current experimental
limit. We have examined cases with $M_3$ less than $400\GeV$ and
have found that almost all models generated are excluded by the
higgs mass limit. The upper value set here is to avoid having too
heavy a gluino.

$G_2$-MSSM models are generated by varying the parameters
$\delta$, $P_{eff}$ and $V_X$ in the following ranges:
\begin{eqnarray}
  -10 \leq \delta \leq 0, \quad 60 \leq P_{eff} \leq 84, \quad V_{X, min} \leq V_X
  \leq
  V _{X, max},
\end{eqnarray}
where $V_{X,min}$ and $V_{X,max}$ are functions of other
parameters and can be found in \cite{ABKKS2}. We only consider the
case in which $Q-P=3$, since other cases either give rise to
extremely heavy gravitinos or lead to AdS vacua
\cite{Acharya:2007rc}.
The gauge couplings and Yukawa couplings at GUT scale are
determined to match the low scale values. The RG evolution is
carried out at 1-loop level with the ``match and run" method to
accommodate large scalar masses. Another consequence of heavy
scalar masses is that the higgs bilinear parameter $Z_{eff}$ is
finely tuned to get EWSB breaking with correct $Z$ boson mass.
$\tan\beta$ is predicted in these vacua to be of $\mathcal{O}(1)$
\cite{ABKKS2}.

\section{Distinguishing String-Susy Models from LHC Signatures}

Thus far we have constructed footprints for four (including two
versions of KKLT) string-susy models. The choices here are made
basically to illustrate the results and techniques with limited
computing. As data approaches and more string-susy models are
added, more systematic calculations can be done. We have shown
that footprints of string theories cover limited regions in LHC
signature space, for understandable reasons. We now examine how to
use these footprints to distinguish among string-susy models at
the LHC. There have of course been discussions on how to
distinguish different beyond-the Standard-Model constructions.
There may be a signature which is sensitive to some features in
the spectra and behaves differently for different models. For
example, same-sign(SS) dileptons is a signature which has been
widely discussed in the literature in distinguishing
supersymmetric models from non-supersymmetric ones. However, this
is often not very useful for distinguishing among various
supersymmetric models, particularly those with an underlying
high-scale. This is because the overall mass scale in each of
these models is not fixed implying that the signatures can vary in
a big range, which will wash out some simple correlations between
(counting) signatures and features in spectrum. So it often
happens that for most of the signatures the two scenarios overlap
a lot and one can not tell them apart completely. But as we have
seen there are correlations between certain pairs of signatures
because of the structure of the underlying theory. This implies
that it is likely that overlapping models of two different
string-susy models do not overlap for other signatures. So
systematically one would try to scan combinations of two
signatures and check if the footprints are completely separated in
the corresponding plots. This method was first explored in
\cite{Kane:2006yi} and was found to be useful in some simple
situations where the spectra of different string-susy models have
big differences. For the string-susy models considered in this
paper, the $G_2$ models can be distinguished from those of KKLT-1
and LARGE Volume by this method. For KKLT-1 and LARGE Volume
models, we have found that no single 2D plot, i.e. no pair of
signatures, can distinguish these models completely. However, as
we will see below, these can still be distinguished by a
combination of several 2D plots.

\subsection{Extracting Correlations from 2D Plots}

In this section we consider a more systematic way to extract
correlations from signatures. Intuitively, one keeps track of the
microscopic parameters associated with points in the overlap
regions and eliminates parameters whenever they are not in an
overlap region. In the remainder of this section and the next
subsection we present some more technical and quantitative
procedures. One should not lose sight of the essential point that
one is distinguishing the theories by adding signature plots and
keeping track of the microscopic parameters of the points in
overlap region.

Conventionally one might try distinguishing classes of models by
directly calculating the distance in the multi-dimensional
signature space including a large number of signatures. A
$\chi^2$-like quantity \cite{ArkaniHamed:2005px} could be defined
with $N_{sig}$ signatures as:
\begin{eqnarray}
  (\Delta S_{A_{i}B_{j}})^2=\frac{1}{N_{sig}}\sum_{a=1}^{N_{sig}}\left(\frac{
  s^{A_{i}}_{a}-s^{B_{j}}_{a}}{\sigma^{A_{i}B_{j}}_{a}}\right)^2.\label{Eq:chisq}
\end{eqnarray}
Here $s^{A_i}_{a}$ is signature $a$ of model $A_i$ and similarly
for $B$. The quantity $\sigma^{A_{i}B_{j}}_{a}$ \footnote{The
definitions of $\sigma^{A_{i}B_{j}}_{a}$ and $(\Delta S_{0})^2$
can be found below Eq.(\ref{Eq:DeltaS2})} characterizes the
uncertainty in the $a^{th}$ signature for the classes of models
$A$ and $B$. If the quantity $(\Delta S_{A_{i}B_{j}})^2$ is
greater than the statistical fluctuation $(\Delta S_{0})^2$ for
all $i$'s and $j$'s, then one should be able to distinguish the
two classes of models. However this method is not as effective as
what we are proposing here. The average over a large number of
signatures in Eq.(\ref{Eq:chisq}) will diminish the difference
between two classes of models if most of the signatures included
are not effective in distinguishing them. A pre-selection of
``useful" signatures could help but there is no systematic {\it a
priori} way of knowing that. Indeed, this conventional method
might not be useful at all, while the method we describe below
always is.

\begin{figure}[h!]
\begin{center}
 \epsfig{file=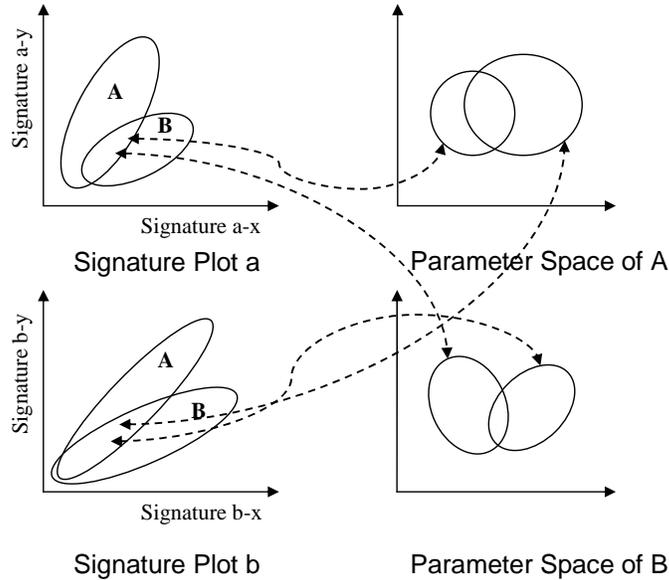,height=8cm, angle=0}
\end{center}
\caption{Figure illustrating the idea that correlation between
different signature plots can be used to reduce ``false degeneracy".}
\label{fig:example}
\end{figure}

Let us start with the following toy example. Suppose there are two
signature plots $a$ and $b$, each of which partially distinguishes
footprint $A$ and $B$. In other words, there is a sizable overlap
between $A$ and $B$ for each plot, denoted as $(A\cap B)_a$ and
$(A\cap B)_b$ respectively. If the signatures in plot $a$ are
correlated non-trivially with signatures in plot $b$, one would
expect that at least some of the models of footprint $A$ in the
overlap $(A\cap B)_a$ can be differentiated from footprint $B$ by
signatures in the plot $b$, and so the set of models of footprint
$A$ in the overlap $(A\cap B)_a$ will have a smaller intersection
with footprint $B$ in $(A\cap B)_b$. In other words, $(A\cap
B)_a\cap(A\cap B)_b$ is smaller than either $(A\cap B)_a$ or
$(A\cap B)_b$. Therefore, the overlap region in 2D signature plots
$a$ or $b$ does not imply a real degeneracy as it is lifted (at
least partially) when more signatures are included. This idea is
illustrated in Fig.\ref{fig:example}. In principle one can
continue adding more signature plots and the overlap region is
expected to be significantly reduced in the end.
To technically realize the above idea, we need to make a few
definitions to quantify the overlap of footprints in the
following. First of all, we define the notion of degeneracy for
two points $A_i\in A$ and $B_j\in B$:

{\it Defn: Two points $A_i\in A$ and $B_j\in B$ are said to be
degenerate in the 2D signature space $(x,y)$ if the $\chi^2$-like
quantity $(\Delta S_{A_{i}B_{j}})^2$ of the two points is smaller than
the statistical fluctuation $(\Delta S_{0})^2$.}

For two signatures, $(\Delta S)^2$ in Eq.(\ref{Eq:chisq}) becomes:
\begin{equation}\label{Eq:DeltaS2}
  (\Delta S_{A_{i}B_{j}})^2=\frac{1}{2}\left[\left(\frac{
  s^{A_{i}}_{x}-s^{B_{j}}_{x}}{\sigma^{A_{i}B_{j}}_{x}}\right)^2
+\left(\frac{s^{A_{i}}_{y}-s^{B_{j}}_{y}}{\sigma^{A_{i}B_{j}}_{y}}\right)^2\right],
\end{equation}
which characterizes the distance in the 2D signature space. Here
$s^{A_i}_{x}$ is signature $x$ of model $A_i$ and similarly for
others. The variance $(\sigma_{x}^{A_{i}B_{j}})$ is defined as
$(\sigma_{x}^{A_{i}B_{j}})^2=(\delta s_{x}^{A_{i}})^2+(\delta
s_{x}^{B_{j}})^2+\Big(f_i(s_{x}^{A_{i}}+s_{x}^{B_{j}})/2\Big)^2$,
where $f_i=0.01$ for all counting signatures and $\delta
s=\sqrt{s+1}$ for counting signatures \cite{ArkaniHamed:2005px}.
$\Delta S_{0}$ characterizes the statistical error, which is
determined by simulating a large number of models with different
random number seeds and taking the $95^{th}$ percentile of the
$\Delta S$'s.

{\it Defn: The model $A_i$ is said to be degenerate with the
entire footprint $B$ with respect to the 2D signature plot $(x,y)$
if there exists at least a model $B_j\in B$ such that $A_i$ and
$B_j$ are degenerate.}
\begin{figure}
  \begin{eqnarray}
  \begin{array}{cc}
  \includegraphics[width=320pt]{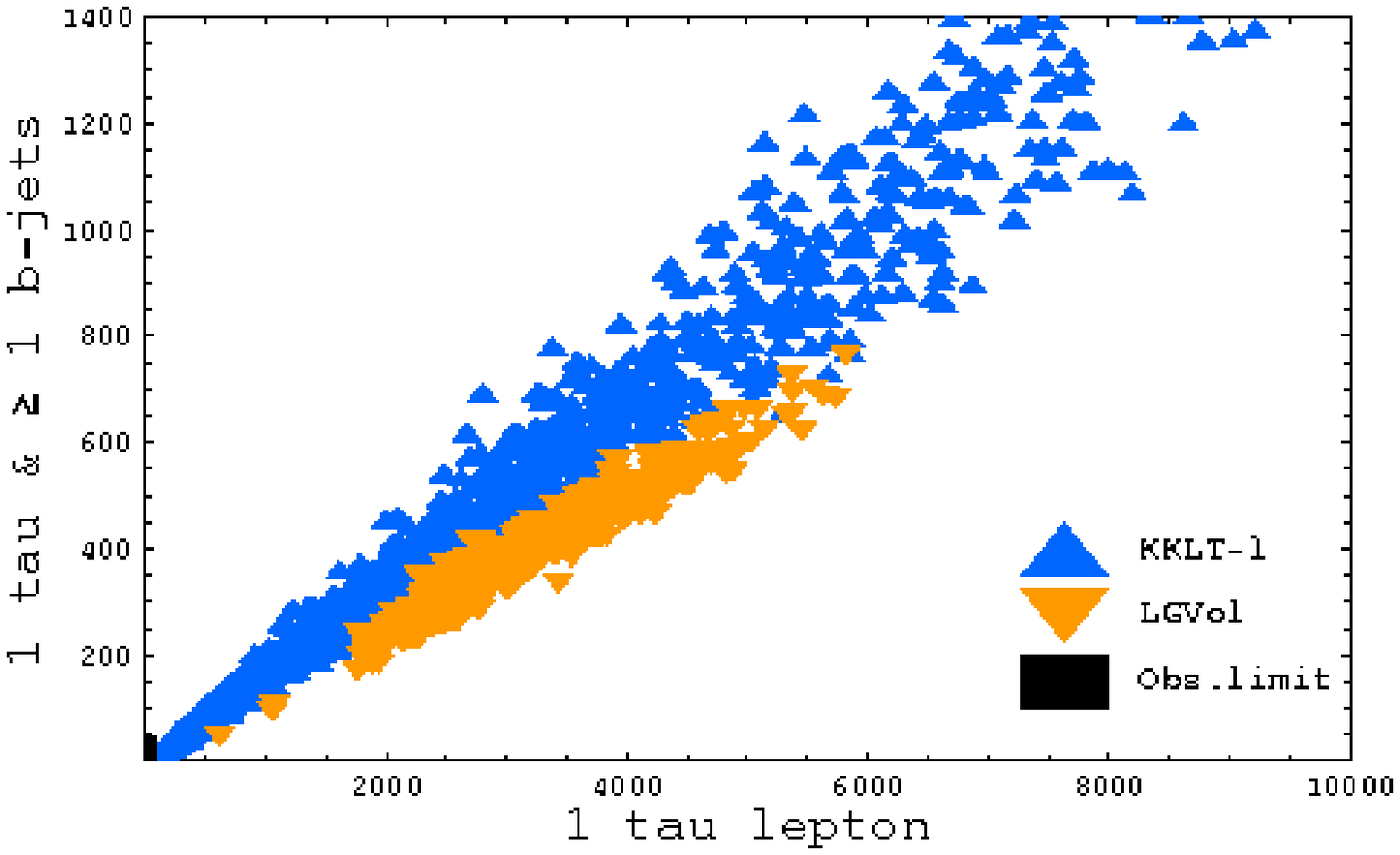}\\
  \includegraphics[width=320pt]{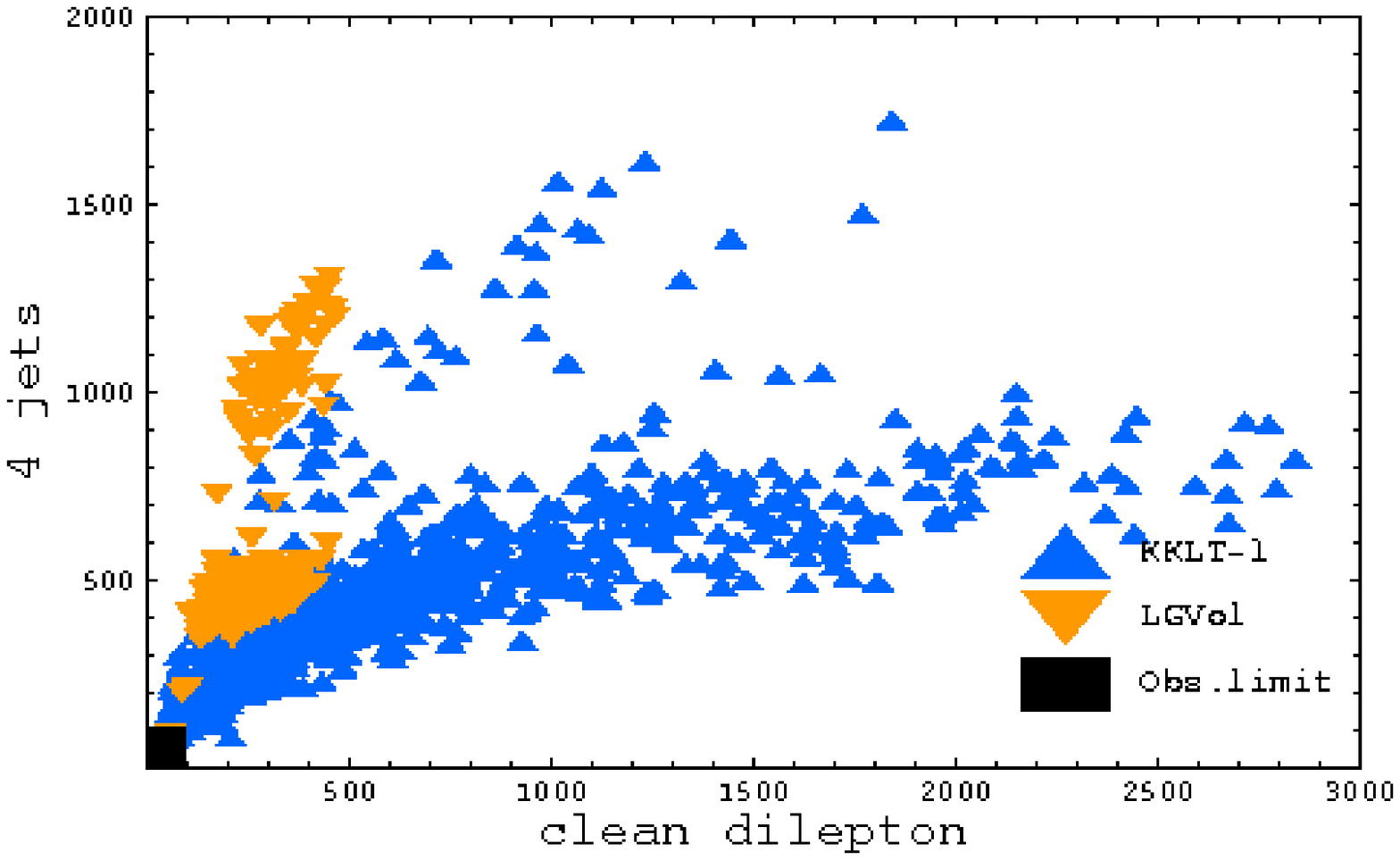} \\
  \includegraphics[width=320pt]{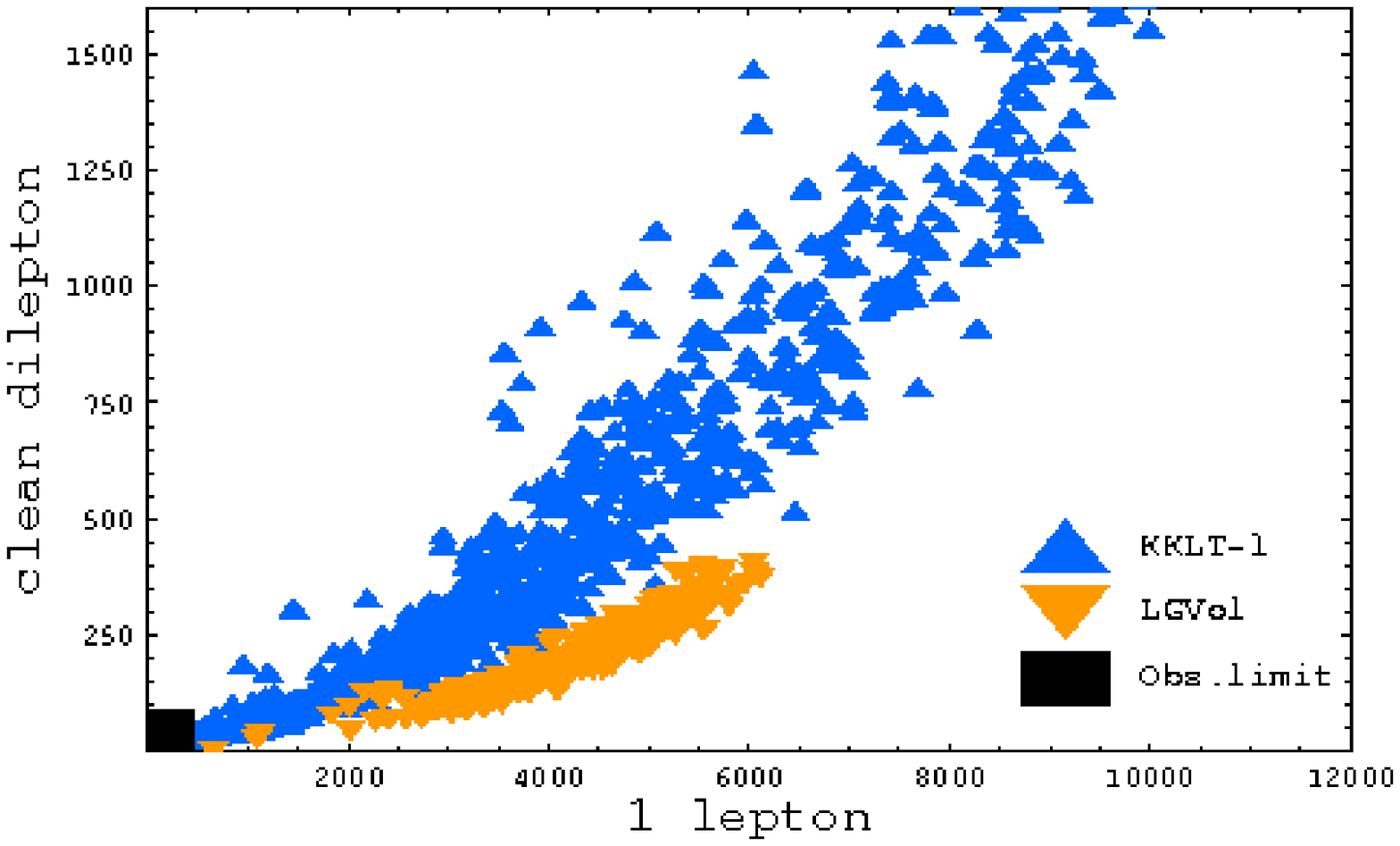}
 \end{array}\nonumber
  \end{eqnarray}
  \caption{Signature plots used for eliminating degeneracy between KKLT-1 (blue)
  and LGVol (orange) string-susy models. All models are simulated with 5$fb^{-1}$ luminosity in PGS4
  with L2 trigger. For the signature - ``1 b-jet and 2 leptons'' and ``clean dilepton",
  there is no requirement of two hard jets.
  For all other signatures, there are at least two hard jets and large missing transverse energy.
  For each example, the points are generated by varying the microscopic parameters over their full ranges,
  as explained in Section \ref{Sec:scan}.}\label{fig:plots2}
\end{figure}

In our convention the number of models of footprint $A$ which are
degenerate with $B$ is denoted by $N_{A,B}$, similarly the number
of models of footprint $B$ which are degenerate with $A$ is
denoted by $N_{B,A}$. One can also notice that $N_{A,B}$ and
$N_{B,A}$ are in general different, because the densities of
models of the two footprints in the overlap region are different
in general. So the overlap can be characterized by the algebraic
mean $N(A,B)\equiv\frac{1}{2}(N_{A,B}+N_{B,A})$. It is clear from
this definition that if the overlap calculated for footprints $A$
and $B$ vanishes, then the corresponding constructions can be
completely distinguished.

As an example, we use the above method to distinguish (original)
KKLT and LARGE Volume string-susy models. To estimate $\Delta
S_0$, we have resimulated 100 KKLT models with different random
numbers and calculated $\Delta S$ for them. The $95^{th}$
percentile of the $\Delta S$ distribution gives $\Delta S_0\approx
1.5$. For different pairs of signatures $\Delta S_0$ will vary by
$\sim \pm 0.1$. As we have mentioned before, the first step of our
strategy is to construct a large set of signature pairs. Without
knowing which ones are better in distinguishing the two classes of
models, one has to add all of them one by one. However to make the
demonstration simple, we first do some trial-and-error analysis
and find some signature plots which can partially distinguish
these two classes of models. This will make the overlap decrease
faster. For the present case, it is actually not difficult to find
these if the $P_{T}$ cut for jets is increased to $200\GeV$ based
on the features in spectrum as explained in \ref{enu:features}.
Three of these plots are shown in Fig. \ref{fig:plots2}. First we
consider those models of KKLT-1 obtained with scan choice 1
(explained in (\ref{choices})) and use the three plots in Figure
\ref{fig:plots2} in our analysis:
\begin{itemize}
\item{1 tau lepton vs. 1 tau lepton and $\ge$ 1 b-jets.}
\item{clean dilepton\footnote{Clean dilepton signature is defined
as the number of dilepton events with no hard jets passing the
event selection cuts.} vs. 4 jets.} \item{1 lepton vs. clean
dileptons.}
\end{itemize}
Starting from the first plot, distances $\Delta S_{A_{i}B_{j}}$
between models in the two classes are calculated and those models
of each class in the overlap region are selected. Then for the
second plot the same procedure is performed except that those
selected models in the previous plot are used instead. After that
we will have a selection of models for each class which still
remain in the overlap region for both plots. This procedure is
carried out by adding more 2D signature plots until either the
number of models in the intersection vanishes or does not decrease
further. For the three plots in the order they are listed, we find
the number of models in each class remained after each operation
decreases monotonically as follows:
\begin{eqnarray}
  \textrm {KKLT-1 (scan choice 1):} &&\quad 119\rightarrow 4\rightarrow 0\nonumber\\
  \textrm {LARGE Volume:}&&\quad 237\rightarrow 17 \rightarrow 0.\label{sequence1}
\end{eqnarray}
To test the stability of this sequence upon changes in $\Delta
S_0$, we use $\Delta S_0=1.7$ and find a similar sequence
\begin{eqnarray}
  \textrm {KKLT-1 (scan choice 1):}&&\quad 129\rightarrow 5\rightarrow 0\nonumber\\
  \textrm {LARGE Volume:}&&\quad 259\rightarrow 21 \rightarrow 0.
\end{eqnarray}
One can see that the number of models in the intersection quickly
decreases as more plots are included. The final overlap of the two
string-susy models is zero, which indicates that they can be
distinguished readily at the LHC even with low luminosities.
Furthermore, the models in the overlap region of each plot as well
as the intersections of these overlap regions can be mapped back
to the parameter space of the underlying string-susy model.

The exact number of models in the final overlap depends on how the
parameter space is scanned and also how densely it is scanned. To
make the method statistically robust, one should sample the
parameter space with a large enough number of points. Furthermore,
to have a reliable count of models in the overlap region, the
density of the points in the footprint should be large enough (at
least for one class of models). In order to confirm that our
result obtained with a sample of 500 points for these classes of
models is robust\footnote{Of course, for different models, one
would need to sample a different number of points in general
depending on the structure of the microscopic parameter space.},
we include 1000 more points for the KKLT string-susy models
(corresponding to scan choices 2 and 3)\footnote{KKLT-1 models
with these two choices have very similar footprints in the
signature space and so including them will increase the footprint
density significantly.}, and try to construct the sequence
(\ref{sequence1}) again. We find the following:
\begin{eqnarray}
  \textrm {KKLT-1 (All scan choices):}&&\quad 451\rightarrow 37\rightarrow 6\nonumber\\
  \textrm {LARGE Volume:}&&\quad 477\rightarrow 289 \rightarrow
  69.\label{sequence2}
\end{eqnarray}
Now the number of models in the overlap does not vanish as before.
However, when we consider three different combinations of the same
signatures as that used earlier, namely:
\begin{itemize}
\item{1 lepton vs. 1 tau lepton} \item{1 lepton vs. 4 jets}
\item{1 lepton vs. 1 tau lepton and $\ge$ 1 b-jets}
\end{itemize}
in addition to the previous combinations, the sequence again
converge to zero as follows
\begin{eqnarray}
  \textrm {KKLT-1 (All scan choices):}&&\quad 451\rightarrow 37\rightarrow 6\rightarrow
  4 \rightarrow 1 \rightarrow 0\nonumber\\
  \textrm {LARGE Volume:}&&\quad 477\rightarrow 289 \rightarrow 69\rightarrow
  11 \rightarrow 1 \rightarrow 0.\label{sequence21}
\end{eqnarray}
For $\Delta S_0=1.7$, we have
\begin{eqnarray}
  \textrm {KKLT-1 (All scan choices):}&&\quad 506\rightarrow 49\rightarrow 10\rightarrow
  8 \rightarrow 7 \rightarrow 4\nonumber\\
  \textrm {LARGE Volume:}&&\quad 488\rightarrow 331 \rightarrow 114\rightarrow
  56 \rightarrow 18 \rightarrow 5,
\end{eqnarray}
which is almost as good as the sequence (\ref{sequence21}). We
learn from the above that the overlap $N(A,B)$ can increase with a
denser scan of parameters. However if two classes of models can be
distinguished intrinsically then $N(A,B)$ will eventually vanish
as more signature plots are included. If one finds $N(A,B)$
approaches a nonzero value even when all possible combinations of
signatures are included (for a given luminosity), then the two
classes of models can not be distinguished completely. We will see
in the next section that it is possible to define a quantity
(which is independent of $N(A,B)$) to characterize the extent to
which two classes of models can be distinguished.

In the above examples, the (close to) optimal set of useful
signatures was arrived at by a judicious use of the
trial-and-error method, i.e. by trying various signature plots
sensibly based on the qualitative features of the classes of
models described above. This procedure should work for other
classes of models as well. However the main purpose of doing this
here is to illustrate the idea without making it too complicated.
In practice, a more systematic way to distinguish classes of
models and pick out useful signatures is to simply add all kinds
of possible signatures and keep track of the overlap. A sharp
decrease in the overlap usually indicates that the signature pair
just added is ``good". In doing this, one does not need to know
much about the features in the models and how they are related to
the signatures, and so the procedure can be implemented in an
automatic way. In future studies, the above procedure could be
supplemented with modern statistical techniques such as neural
networks, boosted decision trees, etc.

\subsection{A Quantitative Definition of Distinguishibility}

In this subsection, we propose a quantitative way to characterize
the distinguishibility of two string-susy models. Let us denote
the two classes of models as $A$ and $B$. Suppose we can properly
define a metric on signature space and hence the volume of the
overlapping submanifold. Then a proper definition of the
distinguishibility could be something like:
\begin{equation}
\eta(A,B)=1-\frac{S(A\cap B)}{2S(A)}-\frac{S(A\cap
B)}{2S(B)},\label{volume}
\end{equation}
where $S(A)$ denotes the volume occupied by $A$ in signature space and similarly for others.
From this definition, we can see $0\leq \eta(A,B) \leq 1$. Clearly
if there is no overlap, i.e. $S(A\cap B)=0$, $\eta(A,B)$ is equal
to 1 indicating that the two footprints can be distinguished completely.
On the other hand, if the two footprints completely
overlap with each other, i.e. $S(A\cap B)=S(A)=S(B)$, then
$\eta(A,B)$ is zero indicating that they can not be distinguished
at all. For other intermediate cases, $\eta(A,B)$ is between $0$
and $1$ indicating partial distinguishibility, which is still
useful depending on the location of experimental data. This will be
discussed in detail in the next section.

In practice,the footprint is sampled by a large number of points instead of being a smooth
manifold. For simplicty we will first assume that the points we sample
are evenly distributed in the signature space, or more
precisely, the number of point in a given region is proportional to
its volume. This is not a realistic assumption, and we will relax it soon.
In this case, the definition of $\eta$ can be rewritten as:
\begin{eqnarray}
  \eta(A,B)=\lim_{N_A, N_B\rightarrow \infty}\left( 1-\frac{N_{A,B}}{2N_A}-
\frac{N_{B,A}}{2N_B}\right),\label{distinguish-def}
\end{eqnarray}
where $N_A$ and $N_B$ are the total number of models in footprint
$A$ and $B$ respectively, while $N_{A,B}$ and $N_{B,A}$ are as
defined in the previous subsection. This gives a practical
definition for distinguishibility. One can see that it is the {\it
ratio} of the number of models in the overlapping region and the
total number of models for a given class of models which
contributes to its distinguishibility with another class of
models. In practice, as long as $N_A$ and $N_B$ are large enough,
the $\eta$ value obtained will be very close to the formal value
obtained after taking the limit to $\infty$. For example, for
KKLT-1 and LARGE Volume (with $\epsilon_0=0.2$), we found that
none of the models in the two classes are degenerate. Thus, using
this definition we get $\eta\rightarrow1$, suggesting that the two
string-susy models can be distinguished quite well.

In generic and more realistic cases the distribution of the models
in the signature space is not flat. We now show that the
definition (\ref{distinguish-def}) is a more natural one to use
compared to definition (\ref{volume}) in such cases. It is
reasonable to assume that sample points are generated uniformly in
model parameter space. If we take into account the mapping to
signature space, the points which sample the footprint are
therefore not likely to be evenly distributed, but are rather
assigned with a probability which is determined by the non-trivial
mapping function. Suppose in the overlap region of two footprints
this probability is small, then the number of points in that
region is also comparatively small relative to the total number of
points even though the volume of the overlap region is not. From
definition (\ref{volume}), $\eta$ and hence the distinguishibility
would be small. However, using definition (\ref{distinguish-def}),
$\eta$ would and hence the distinguishibility would be large. This
seems more natural since by assumption, it is much less likely to
populate the overlap region compared to the non-overlap region by
scanning model-parameters. Therefore we see that the definition
(\ref{distinguish-def}) in terms of the number of points is quite
reasonable and convenient in practice.

\section{Discussion and Conclusion}

In this paper, we have introduced the idea of constructing
footprints of ``string-susy models" (defined in Section IIA), and
a general technique to distinguish different models by
correlations of signatures. Focusing on four classes of
string-susy models where calculations are reliable, our first
major result is that they all give limited footprints in signature
space. In addition, the LHC signatures of a particular class of
models are sensitive to at least some of the underlying
theoretical structure. This information is not only encoded in the
values of signatures themselves but also in their correlations.
Familiar inclusive dilepton and trilepton signatures are not very
helpful in distinguishing among these string-susy models. However
they can be distinguished by systematically adding and studying
the pattern of signature plots and qualitatively understanding
their origin. We have explicitly shown that the overlap area of
two footprints becomes smaller and finally vanishes as more
signatures plots are included.
\begin{figure}
\leavevmode \epsfxsize 12 cm \epsfbox{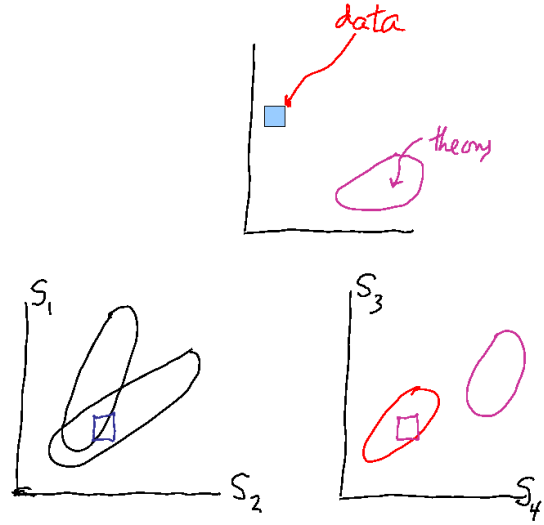} \caption{Cartoon to
illustrate the various possibilities once there is data. The
square box denotes ``data'' while the elliptical regions denote
slices of footprints of two string-susy models in two signature
plots.}\label{cartoon}
\end{figure}

Of course, it may be possible to recognize and interpret what is
discovered rather easily. Our approach will worthwhile
particularly even when superpartner masses and properties can not
be untangled, and when degeneracies are present, so more
traditional approaches work poorly. These methods can be applied
with very limited data, and improved as integrated luminosities
increase and more signatures become available.

The construction of footprints will of course be especially useful
when the LHC data is available, which will appear as a box in the
signature space. Suppose there are two string-susy models which
can be distinguished from each other using the method described in
this paper, with low luminosity data at the LHC. In the event of
actual data, if the box corresponding to the data is far away from
the footprints of both models, then both string-susy models are
excluded. If the box is inside one of the footprints but not the
other one, then the corresponding string-susy model is favored
while the other is excluded. However, if the box lies in the
overlap region of two footprints\footnote{That is the box lies in
the overlap of the entire footprints, not just some 2D
footprints.}, this means that both string-susy models are
consistent with data at that particular luminosity. The cartoon in
figure \ref{cartoon} illustrates the point. In this case, one can
focus on all such models and both improve the theoretical
constructions and add experimental observables and associated
signature plots to distinguish among them further.

We have focussed on LHC counting signatures to demonstrate the
methods. Other LHC data such as asymmetries (Figure \ref{asymm}),
distributions in $\notEt$, $p_T$, mass pairs, etc. all have
limited footprints as well, and can be valuable. Dark matter
detection and relic density involve the same physics and are
meaningful to include for the signatures of an underlying
microscopic theory (but not for a low-scale effective theory), as
is $g_{\mu}-2$. Adding these signatures will make the approach
even more powerful. Since one has a full theory, it is meaningful
to add them.

In the steps from high-scale string construction to LHC
signatures, there are various kinds of uncertainties which were
not yet considered properly. These include uncertainties in the
ranges of the high-scale parameter space, uncertainties arising
from the RG evolution codes, uncertainties arising from those in
the SM parameters (the top quark mass for example), uncertainties
in experimental constraints, as well as those arising from
simulations - event generation, parton showering and hadronization
and detector simulation. One might worry whether these
uncertainties will change the effectiveness of our method. Of
course, the boundary of the footprint will become fuzzy when these
uncertainties are included. However since our analysis is based on
the correlations of signatures, the distinguishibility of any two
scenarios should not be affected. A better investigation of these
issues is left for the future, as is the study of theories that
give extended MSSMs for the visible sector. We expect the study of
patterns of signature plots to be equally valuable in such cases.

Although particular classes of string-susy models with the same
visible sector - the MSSM - can be distinguished, it may be more
challenging to distinguish classes of string vacua with different
matter and gauge spectra. If it turns out that there are some
exotic fields beyond the MSSM which are light enough to be
produced (on-shell or off-shell), there could be a substantial
change in LHC signatures.

For all the string-susy models considered here, Electroweak
symmetry breaking (EWSB) is accommodated but not predicted, in the
sense that the $\mu$ parameter is chosen by hand so as to satisfy
the EWSB condition. If one could find a natural dynamical
mechanism giving EWSB in the future, it would be important to know
what the effect of this natural mechanism would be on the pattern
of signatures computed by just accommodating EWSB. The answer to
this question is not definitive, since it depends on the particle
spectrum and the exact mechanism. For example, within the context
of the MSSM, it is well known that consistent EWSB requires a
precise relation between the supersymmetric $\mu$ parameter and
some soft supersymmetry breaking parameters. Therefore, one
natural solution to the problem of EWSB is that this precise
relation can be predicted or explained from the structure of a
string-susy model. In this case, there would be no effect on the
pattern of signatures. However, if the visible sector of the
string-susy model consists of particles in addition to the MSSM
which are light enough to be produced at the LHC and play a
non-trivial role in the EWSB mechanism (additional $U(1)$ models
are an example), then the pattern of signatures could be affected
significantly. Nevertheless, the methods described in the paper
are still applicable.

We have proposed a new approach to relating collider data and an
underlying theory. In the case where the underlying theory is a
string construction with stabilized moduli and softly broken
supersymmetry, we have seen that particular constructions give
very limited footprints in signature space, and that LHC
signatures for a particular class of vacua are sensitive to at
least some of the underlying structure of the theory. Not all
theories agree with data, and the subset that do can be
distinguished by considering the pattern of a number of
signatures. The software techniques needed to carry out such a
program for a variety of constructions mostly already exist, and
are improving. Our analysis needs to be extended in a number of
directions, perhaps most by examining a number of constructions
from different corners of string/$M$ theory, with different
compactifications, different ways of generating de Sitter vacua
and different ways of breaking supersymmetry in a controlled
manner.

We think that string theorists will learn about string theory by
studying collider phenomenology. This has happened from studying
the visible sector, particularly in heterotic and Type II toroidal
constructions, and we expect it to happen from the study of
superpartner properties. Given the large number of string vacua
one could ask whether it is very unlikely that the ones we study
could be like our vacuum? We think it is not so unlikely because
we do not study random string/$M$ theory constructions - we select
for study those that can give SM-like matter, softly broken
$\mathcal{N}=1$ supersymmetry, dark matter, inflation, and so on.
The approach described here may help us learn if we live in a
string/$M$ theory vacuum, and learn more about its properties.

\acknowledgments The authors appreciate encouragement and helpful
conversations with and suggestions from Bobby Acharya, Joseph
Lykken, Arjun Menon, David Morrissey, Brent Nelson, Aaron Pierce,
Albert de Roeck, Maria Spiropulu and Liantao Wang. The research of
GLK, PK and JS is supported in part by the US Department of
Energy.

\section{Appendix: Counting Signatures used in our Study}

As we have discussed before, we are particularly interested in
extracting information from the low luminosity data (5-10
$fb^{-1}$) corresponding to 1-2 year running of LHC. For this
purpose, we will select a special set of counting signatures as
our observables. A complete set of counting signatures can be
found in \cite{ArkaniHamed:2005px}. Here we consider the following
counting signatures:
\begin{itemize}
\item{ 1 lepton, OS dilepton, SS dilepton, trilepton, 1 tau
lepton, 2 tau leptons, 3 tau leptons, OSSF dilepton, OSDF
dilepton, SSSF dilepton, SSDF dilepton, OS dilepton(e,$\mu$), SS
dilepton(e,$\mu$)} \item{ 1 jet, 2 jets, 3 jets, 4 jets, 1 b-jet,
2 b-jets, 3 b-jets, 4 b-jets}\item{2 leptons and 1 jet, 2 leptons
and 2 jets, 2 leptons and 3 jets, 2 leptons and 4 jets}\item{0
lepton and 2 b-jets, 0 lepton and $\geq$ 3 b-jets, 1 lepton and
$\geq$ 2 b-jets, 2 leptons and 0 b-jet, 2 leptons and 1 b-jet, 2
leptons and 2 b-jets, 2 leptons and $\geq$ 3 b-jets, 3 leptons and
1 b-jet} \item{1 tau and $\geq$ 1 b-jets, 1 tau and $\geq$ 2
b-jets, 2 tau and $\geq$ 2 b-jets, $\geq 2$ tau and 1 b-jet}
\item{1 positive lepton, 1 negative lepton, clean dilepton}
\end{itemize}

\end{document}